\documentclass[a4paper,11pt]{article}
\pdfoutput=1 

\usepackage{jheppub} 

\usepackage{multirow}
\usepackage{tabularx}
\usepackage{float}

\title{Anatomy of kaon decays and prospects for lepton flavour universality violation}

\author[a]{G.~D’Ambrosio,}
\author[b]{A.M.~Iyer,}
\author[c,d]{F.~Mahmoudi}
\author[a]{and S.~Neshatpour}

\affiliation[a]{INFN-Sezione di Napoli, Complesso Universitario di Monte S. Angelo, Via Cintia Edificio 6, 80126 Napoli, Italy}
\affiliation[b]{Department of Physics, Indian Institute of Technology, Delhi, Hauz Khas, New Delhi-110016, India}
\affiliation[c]{Universit\'e de Lyon, Universit\'e Claude Bernard Lyon 1, CNRS/IN2P3, 
Institut de Physique des 2 Infinis de Lyon, UMR 5822, F-69622, Villeurbanne, France}
\affiliation[d]{Theoretical Physics Department, CERN, CH-1211 Geneva 23, Switzerland}

\emailAdd{gdambros@na.infn.it}
\emailAdd{iyerabhishek@physics.iitd.ac.in}
\emailAdd{nazila@cern.ch}
\emailAdd{neshatpour@na.infn.it}

\abstract{The kaon sector is characterised by several processes which are under active investigation across different experiments. In this work,
we present the  global picture that emerges from a study of the different decay modes. We begin by revisiting the theoretical component of these decays and providing up-to-date predictions of the Standard Model as well as the corresponding uncertainties. Several new features emerge, in particular for $K_{S,L}\to \mu\bar\mu$, and are presented in considerable detail. This offers an ideal platform for extracting the parameter space supported by the existing data.
Motivated by possible lepton flavour universality violation in $B$~decays, we investigate such Beyond the Standard Model effects also in rare kaon decays.
Without loss of generality, our primary analyses correspond to the paradigm where the Wilson coefficients  for operators involving tau leptons are chosen to be equal to that involving the muon, i.e.~$\delta C^\tau=\delta C^\mu$. We conclude by presenting the possible picture that can be achieved towards the end of the run of data accumulation  in the planned experiments. This includes assumptions on possible sensitivity goals that the experiments can aim to achieve, in order to extract the kind of physics highlighted in this~paper.}

\arxivnumber{2206.14748}

\begin{document} 
\maketitle
\flushbottom

\vspace{-0.5cm}
\section{Introduction}
The last decade has seen a period of intense scrutiny for the flavour  structure of the Standard Model (SM). The plethora of experimental results has served as a catalyst to investigate the different sectors, both from the SM point of view as well as looking at possible effects of New Physics (NP) on them. Some of the different measurements which drive the interest in the  theoretical community include the anomalous magnetic moment of the muon\cite{Abi:2021gix,Bennett:2006fi}, decays corresponding to $b\to s$ transitions ($B$~physics) \cite{LHCb:2014vgu,LHCb:2017avl,Belle:2019oag,LHCb:2019hip,LHCb:2021trn}, $b\to c$ transitions \cite{BaBar:2013mob,Belle:2015qfa,LHCb:2015gmp,LHCb:2017smo,Belle:2019gij,LHCb:2022piu}, nuclear beta decays and implications for $V_{ud}$ \cite{Hardy:2020qwl}, as well as decays corresponding to $s \to d$ transitions (kaon physics) \cite{NA62:2021zjw,Bician:2020ukv,Ahn:2018mvc}. Many of these experiments have hinted at the possibility of non-standard physics in their respective data analysed thus far. 
Among the most interesting hints of New Physics is the observation of lepton flavour universality violation (LFUV) in rare $B$~decays \cite{LHCb:2014vgu,LHCb:2017avl,Belle:2019oag,LHCb:2019hip,LHCb:2021trn}. Global fits using the effective field theory approach proved useful in constructing the appropriate beyond the SM scenarios to explain  LFUV~\cite{Hurth:2021nsi,Alguero:2021anc,Altmannshofer:2021qrr,Ciuchini:2020gvn,Geng:2021nhg,Datta:2019zca,Alok:2019ufo,Kowalska:2019ley,DAmico:2017mtc}.
It is natural to expect that these NP effects would also  impact operators contributing to kaon decays. This provides a strong motivation to consider an effective theory fit dedicated to kaons and to possibly achieve a similar sensitivity to the corresponding analyses in $B$~decays.

There are several observables in $K$~systems that have the capability to make an individual impact on the eventual global fits.
Recently, there has been a significant (and ongoing) experimental effort focused on the measurements of the branching fractions of $K^+\to \pi^+ \nu \bar{\nu}$ (NA62 at CERN~\cite{NA62:2021zjw}) and $K_L \to \pi^0 \nu \bar{\nu}$ (KOTO at J-PARC~\cite{Ahn:2018mvc}).  Both of them have negligible uncertainty on long-distance contributions, making them highly sensitive to non-standard physics. However, any analysis involving only these decays proves inadequate to make concrete claims about LFUV effects in kaons, thereby prompting the addition of new observables.  In the context of direct sensitivity to LFUV, $K^+\to\pi^+\ell\bar{\ell}$ offers an exciting prospect. It has been shown that the difference of the leading order polynomial expansion coefficients  of the vector form factor in these decays is sensitive to short-distance flavour violating effects \cite{Crivellin:2016vjc}.
The parameter space that is permitted by these three observables can be further limited by the consideration of
decays like  $K_{L,S}\to \ell \bar{\ell}$ and $K_L\to \pi^0\ell\bar{\ell}$.
With the exception of $K_L\to\mu\bar{\mu}$, the others have only upper bounds with respect to their experimental status. While they are weak at present, they are still useful in drawing attention to a specific part of the parameter space of the NP Wilson coefficients.
Each of them ($K_{L,S}\to \ell \bar{\ell}$, $K_L\to \pi^0\ell\bar{\ell}$) is characterised by dominant long-distance effects, making it relatively more challenging to extract non-standard physics. This translates into a limitation on the existing accuracy in their SM computation. However, there exists a well-defined experimental program for each of these decays. This may either imply an improvement in the existing sensitivities  or a measurement at the SM level and is outlined in the third column of table~\ref{tab:data}.

In light of several ongoing and planned measurements for decays in the kaon sector, this paper intends to demonstrate the complementary  capability for LFUV measurements in kaon systems. We analyse each of these decays and make a careful evaluation of the theoretical uncertainties. Using the updated values from CKM and other input parameters, the uncertainties are computed using a Monte Carlo approach. An interesting difference from past literature is observed  for the $K_L\to \mu\bar{\mu}$ decay which is described by asymmetric uncertainties. Integrating these decays in {\tt{SuperIso}}~\cite{Mahmoudi:2007vz,Mahmoudi:2008tp,Mahmoudi:2009zz,Neshatpour:2021nbn,Neshatpour:2022fak}, the relevant parameter space of the New Physics Wilson coefficients is identified. Guided by a well-defined strategy for the measurement of many of these decays, the experimental uncertainties are used appropriately. Furthermore, for decays for which no such well-defined strategy exists, we also present  projections on the progress on  the experimental side. Demonstrating a rich yield of interesting physics  could motivate modified strategies for such decays in the future. This is particularly true for the measurement of vector form factors in $K^+\to\pi^+\ell \bar{\ell}$.
While  measurements of these form factors exist for both the electron ~\cite{E865:1999ker,NA482:2009pfe,NA482:2010zrc} and the muon \cite{Bician:2020ukv},  a strong case for higher precision measurements of these quantities is  presented in this work. Similarly, our results present the need for a reduction in the error on the theoretical computation of $K_L\to\mu \bar{\mu}$.

The paper is organised as follows: in section~\ref{sec:2} we analyse the decay modes of interest in considerable detail. The considered processes are $K^+(K_L)\to \pi^+(\pi^0)\nu\bar\nu$ in section~\ref{sec:ktopinunu}, LFUV in $K^+\to\pi^+\ell\ell$ decays in section~\ref{sec:lfuv}, $K_{S,L}\to \mu \bar{\mu}$ in section~\ref{sec:Ksmumu} and $K_L\to\pi^0\ell \bar{\ell}$ in section~\ref{sec:KLtopill}. The analyses in each of these subsections (along with the appendices)  are self-contained and offer an up-to-date evaluation of the SM values as well as the corresponding uncertainties. In section~\ref{sec:global} we present a global picture involving all the decays, which illustrates the existing bounds from the different observables.  
Section~\ref{sec:global1} is devoted to the description of the methodology of our fit. In section~\ref{sec:global2} we perform a global fit to current experimental data. Section~\ref{sec:global3} offers 
possible improvements in the fits at the end of the run for most of the experiments. This includes using the official projections for some observables as well as choosing optimistic reaches for the others. Finally, we conclude in section~\ref{sec:conc}.

\section{Theoretical framework}
\label{sec:2}
In this section, we set up the convention to be followed for the rest of the paper. The $s \to d$ transitions can be parameterised by the following effective Hamiltonian:
\begin{equation}\label{eq:Heff}
\mathcal{H}_{\rm eff}=-\frac{4G_F}{\sqrt{2}}\lambda_t^{sd}\frac{\alpha_e}{4\pi}\sum_k C_k^{\ell}O_k^{\ell}\,,
\end{equation}
where $\lambda_t^{sd}\equiv V^*_{ts}V_{td}$ and the relevant effective operators are
\begin{align}\nonumber
&{O}_9^{\ell} = (\bar{s} \gamma_\mu P_L d)\,(\bar{\ell}\gamma^\mu \ell)\,,
&&{O}_{10}^{\ell} = (\bar{s} \gamma_\mu P_L d)\,(\bar{\ell}\gamma^\mu\gamma_5 \ell)\,,\\
  &{O}_L^{\ell} = (\bar{s} \gamma_\mu P_L d)\,(\bar{\nu}_\ell\,\gamma^\mu(1-\gamma_5)\, \nu_\ell)\,,
\label{eq:operators}
\end{align}
with $P_L=(1-\gamma_5)/2$. The most general Hamiltonian also includes  scalar and pseudoscalar operators, as well as the chirality-flipped counterpart of the above operators where the quark currents are right-handed. In this instance, we focus on this small subset of operators which have the same structure as the most relevant operators for explaining the neutral current $B$-anomalies~\cite{LHCb:2014vgu,LHCb:2017avl,Belle:2019oag,LHCb:2019hip,LHCb:2021trn}.
The Wilson coefficients $C_k^{\ell}$  include any potential (flavour violating) New~Physics contribution parameterised~as\footnote{Within the considered basis, a real $\delta C_i$ results in both real and imaginary  short-distance  contributions in the effective Hamiltonian.}
\begin{equation}
C_k^{\ell} = C_{k,{\rm SM}}^{\ell}+ \delta C_{k}^{\ell}\,.
\end{equation}

In recent years, there has been much progress in the measurements of rare kaon decays.  However, still several of the rare kaon decays have not been observed and there are only upper bounds available for them.  In general, different New Physics contributions with various combinations of the operator structures of eq.~\ref{eq:operators} can contribute to kaon decays. 
Nonetheless, given the rather limited experimental data currently available for rare kaon decays and the fact that New Physics is more conveniently explored in the chiral basis, we limit our analysis to the class of NP scenarios where the charged and neutral leptons are related to each other by the SU(2)$_{\rm L}$ gauge symmetry.
As we consider only left-handed quark currents, the different Wilson coefficients that we consider are related to each other as $\delta C_{L}^{\ell} \equiv \delta C_9^{\ell} = - \delta C_{10}^{\ell}$.
With this background, we set up the theoretical description of the different decay modes in the following subsections.

\subsection{\texorpdfstring{$K^+\to \pi^+\nu \bar{\nu}$}{K+ -> pi+ nu nu} and \texorpdfstring{$K_L\to \pi^0\nu \bar{\nu}$}{KL -> pi0 nu nu}}\label{subsec:Kpinunu}
\label{sec:ktopinunu}
These rare decay modes  receive dominant short-distance (SD) contributions. Their high sensitivity to any  NP effect while having very small theoretical uncertainties justify their status as being among the eagerly awaited measurements from the corresponding experiments \cite{NA62:2021zjw,Ahn:2018mvc}. In the notation discussed above, the 
 branching fractions for these modes are given as~\cite{Bobeth:2017ecx}  (see also~\cite{Rein:1989tr,Hagelin:1989wt,Lu:1994ww,Geng:1995np,Buchalla:1995vs,Fajfer:1996tc,Buchalla:1996fp,Buchalla:1998ba,Mescia:2007kn,Misiak:1999yg,Brod:2010hi})
\begin{align}
  \label{eq:Br-KLpinunu}
  {\rm BR}(K_L \to \pi^0 \nu \bar{\nu}) & =  \frac{\kappa_L }{\lambda^{10}}\frac{1}{3}s_W^4 \sum_{\ell}
  {\rm Im}^2 \left[\lambda_t C_L^{\ell} \right]\,, \\
  \label{eq:Br-Kppipnunu}
  {\rm BR}(K^+ \to \pi^+ \nu \bar{\nu}) & =
  \frac{\kappa_+ (1 + \Delta_{\rm EM})}{\lambda^{10}}\frac{1}{3} s_W^4 \sum_{\ell}
  \left[  {\rm Im}^2 \Big(\lambda_t C_L^{\ell} \Big)
        + {\rm Re}^2 \Big(-\frac{\lambda_c X_{c}}{s_W^2}
                         + \lambda^{sd}_t C_L^{\ell} \Big)\right]\,,
\end{align}
where the sum is over the three neutrino flavours.
Considering the relevant input parameters as collected in appendix~\ref{app:inputs},
we calculate the short-distance SM contribution given by $X_c$
and $C_{L,{\rm SM}}^{\ell} = C_{L,{\rm SM}} ={-X(x_t)}/{ s_W^2}$ (see appendix~\ref{app:Xxt}). 
The values of the branching fraction for the SM, corresponding to these inputs are given in table~\ref{tab:data} where the theory uncertainties are estimated using a Monte Carlo method, assuming Gaussian errors for the input parameters.

\begin{figure}[t!]
\begin{center}
\includegraphics[width=0.48\textwidth]{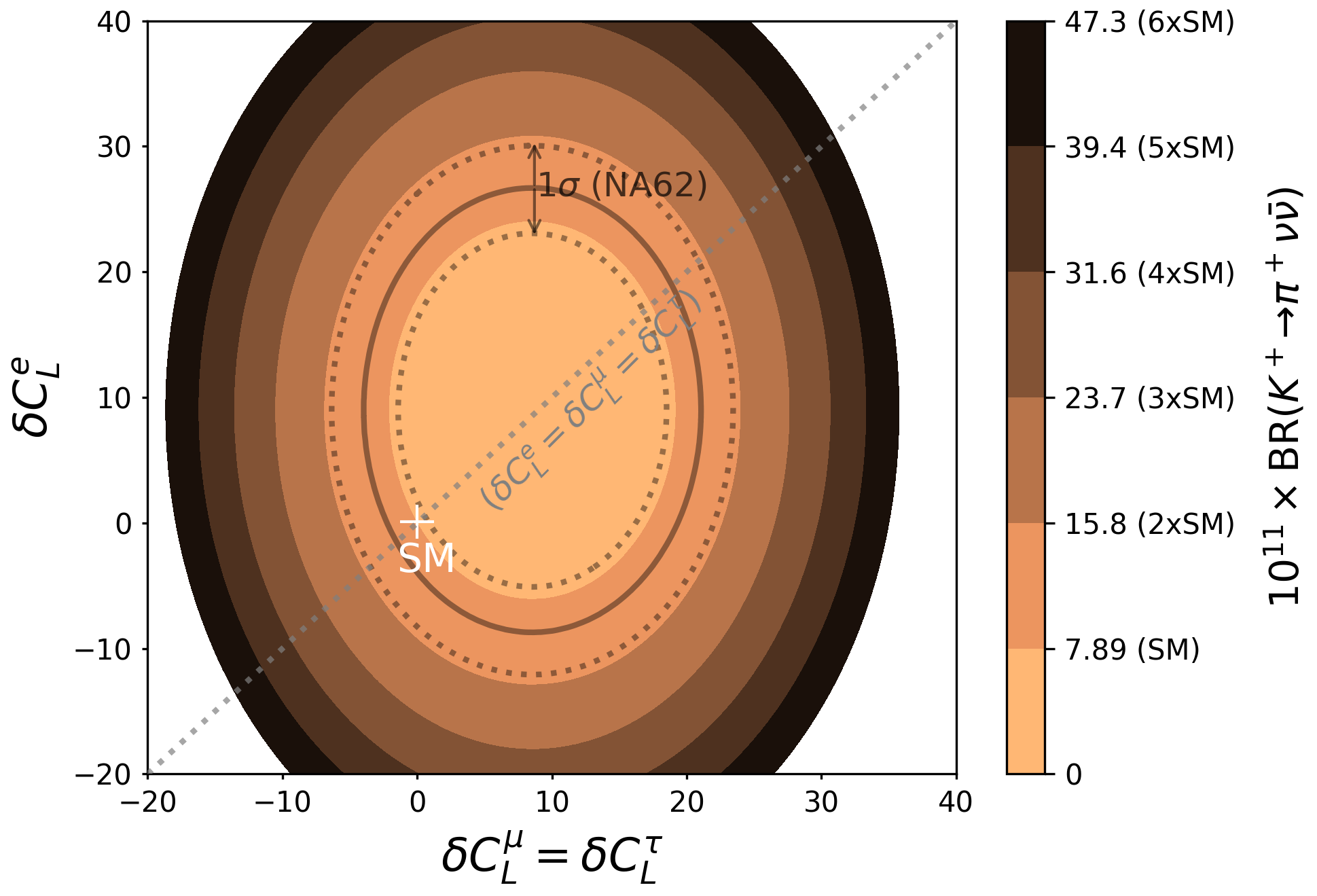}\quad  \includegraphics[width=0.49\textwidth]{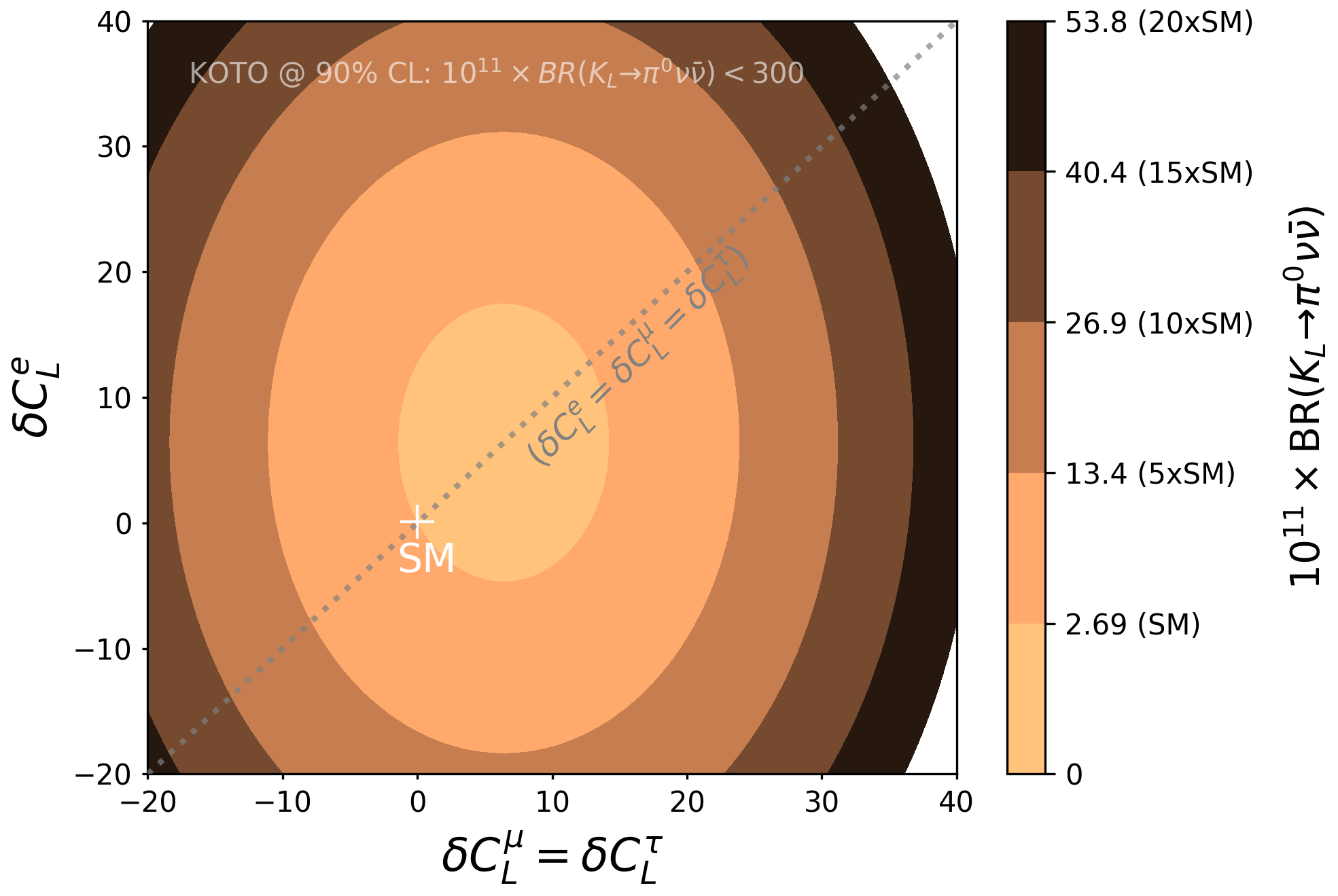}
\caption{\small
BR($K^+\to \pi^+ \bar{\nu}\nu$) (left) and 
BR($K_L\to \pi^0 \bar{\nu}\nu$) (right) as a function of 
$\delta C_L^e$ and $\delta C_L^\mu=\delta C_L^\tau$. The dotted grey line represents the lepton flavour universality scenario.
In the left plot, the brown solid (dotted) line corresponds to the measured central value ($1\sigma$ experimental uncertainty) by NA62~\cite{NA62:2021zjw}. In the right plot, the upper bound on BR($K_L\to \pi^0 \bar{\nu}\nu$) is not visible for the scanned values.
\label{fig:Kppinunu_CLeCLmu}}
\end{center}
\end{figure}

An interesting feature of these  decay modes is that an experimental result consistent with the SM prediction does not necessarily imply the absence of NP. This is due to the fact that summation over the three species of neutrinos can result in a relative cancellation between the corresponding NP Wilson coefficients. This is illustrated in figure~\ref{fig:Kppinunu_CLeCLmu} for  $K^+ \to \pi^+ \nu \bar{\nu}$ (left) and $K_L \to \pi^0 \nu \bar{\nu}$ (right). For simplicity, we have set $\delta C^\tau_L=\delta C^\mu_L$. This facilitates a visual comparison on the departures from the lepton flavour universality  given by the dotted grey line.\footnote{An alternative situation with $\delta C^\tau_L=\delta C^e_L$ is illustrated in appendix~\ref{app:otherpossibility}.}
The figure shows concentric circles, centered at $(\delta C_L^\mu,\delta C_L^e)=(8.5,9.0)$ on the left, and at $(\delta C_L^\mu,\delta C_L^e)=(6.5, 6.5)$ on the right. The steady darkening of the annuli on moving away from the centre represents an increase in the  value of the corresponding branching fraction. At the centre, the contributions due to the SM and the NP Wilson coefficients cancel exactly to result in a null value for the branching fractions. Moving along the circumference of any circle evaluates to the same value of the branching fraction.
In the left plot, the brown solid line represents the current measured value for $K^+ \to \pi^+ \nu \bar{\nu}$ with the corresponding $\pm 1\sigma$ uncertainties given by the dotted lines. In the right plot, the upper bound for $K_L \to \pi^0 \nu \bar{\nu}$ \cite{Ahn:2018mvc} is not visible for the regions scanned in the    $(\delta C_L^\mu,\delta C_L^e)$ plane. 

In this study, we re-estimate the SM predictions for these branching fractions using the updated inputs,  as given in table~\ref{tab:inputs}. They are represented by a cross in figure~\ref{fig:Kppinunu_CLeCLmu}. The corresponding theory uncertainty is not visible on the scale of the figure and the evaluated numbers are quoted below:
\begin{align}
    \text{BR}(K^+\to \pi^+\nu \bar{\nu})^{\rm SM}    &= (7.86 \pm 0.61)\times 10^{-11}\,,\\
    \text{BR}(K_L\to \pi^0\nu \bar{\nu})^{\rm SM}    &= (2.68 \pm 0.30) \times 10^{-11}\,.
\end{align}
We are in agreement with the corresponding evaluation in \cite{Brod:2021hsj}.
 Figure~\ref{fig:Kppinunu_CLeCLmu} also illustrates the fact that a mere observation in agreement with the SM prediction of either of these decays cannot be a conclusive claim for  lepton flavour universality. This is evident by comparing the current measurement for $K^+ \to \pi^+ \nu \bar{\nu}$ with the corresponding SM value as given in table~\ref{tab:data}. The orange band represents the $1\sigma$ region consistent with the current measurement.
Although the SM prediction, which implies flavour universality, is in agreement with experimental measurement, combinations of possibly LFUV NP contributions to $\delta C_L^{e,\mu}$ in the $[-11,29]$ range also result in theoretical predictions within the $1\sigma$ range of the measured value. This prompts the inclusion of other decay modes for the kaons.


\subsection{LFUV in \texorpdfstring{$K^+\to \pi^+\ell \bar{\ell}$}{K+ -> pi+ ll}}
\label{sec:lfuv}

In an attempt to look for observables which may aid in making conclusive observations regarding NP as well as the possibility of lepton flavour universality violation effects, it is natural to look for motivation from $B$~physics.
The $R_{K}$ ratios for testing universality are constructed~\cite{Hiller:2003js} using the $B\to H \ell \bar{\ell}$ processes for $H=(K^{(*)},\phi,...)$. An analogous mode in kaons is the $K^+\to\pi^+\ell \bar{\ell}$. Thus it is natural to explore these modes to construct similar observables in kaon systems.

The branching fractions for $K^+\to\pi^+\ell \bar{\ell}$ decay is dominated by the long-distance contribution $K^+\to\pi^+\gamma^*$ which can be approximated by the following amplitude:

\begin{equation}
A_V^{K^+\to\pi^+\gamma^*}=-\frac{G_f\alpha}{4\pi}V_+(z)\bar u_l(p_-)(\gamma_\mu k^\mu+\gamma_\mu p^\mu)v_l(p_+)\,,
\end{equation}
where $V_+$ is the vector form factor approximated as 
\begin{equation}
V_+(z)=a_++b_+z+V_+^{\pi\pi}(z)\,,
\end{equation}
with $z=\frac{(p_{\ell}+p_{\bar{\ell}})^2}{M_K^2}$ and $V_+^{\pi\pi}(z)$  describing the contribution from the two-pion intermediate state~\cite{DAmbrosio:1998gur}~(see also~\cite{Gilman:1979ud, Ecker:1987hd, Ecker:1987qi, DAmbrosio:1994fgc, Ananthanarayan:2012hu}) with input from the external parameter fit to  $K\to\pi\pi\pi$ data~\cite{Kambor:1991ah,Bijnens:2002vr}, while the parameters $a_+$ and $b_+$ are determined by experiments via a fit to  experimental data on  $K^+\to\pi^+\ell \bar{\ell}$. This can then be used for the SM computations of the corresponding branching fractions~\cite{DAmbrosio:1998gur,Cirigliano:2011ny}.
The assumption of a SM-like pattern while estimating the coefficients $a_+$ and $b_+$ is reasonable on account of  dominant long-distance effects~\cite{ColuccioLeskow:2016noe, DAmbrosio:2018ytt, DAmbrosio:2019xph}.
Thus, any information regarding New Physics contributions due to short-distance physics is hidden and not immediately apparent by noting the individual values of the branching for each channel. 
Nonetheless, a key point  here is that the long-distance effects are purely universal and the same for  all lepton flavours. Thus any deviation from this paradigm is necessarily due to NP contributions. A convenient representation is to take the difference of the coefficients as \cite{Crivellin:2016vjc}
\begin{align}\label{eq:K_LFUV}
 a_+^{\mu\mu}-a_+^{ee} = - \sqrt{2}\,{\rm Re}\left[V_{td}V^*_{ts} (C^{\mu}_{9}-C^{e}_{9}) \right] \,,
\end{align}
where the long-distance part cancels out  and one is only sensitive to the short-distance effects if any. This is also a measure of non-universality between the leptons.

\begin{table}[t]
\renewcommand{\arraystretch}{1.3}
\centering
\scalebox{0.75}{
\begin{tabular}{cccr}
\multicolumn{4}{c}{\emph{Historical progression}}\\\hline\hline
Channel & $a_+$ & $b_+$ & Reference \\
\hline
$ee$ & $-0.587\pm 0.010$ & $-0.655\pm 0.044$ & E865~\cite{E865:1999ker}\\
$ee$ & $-0.578\pm 0.016$ & $-0.779\pm 0.066$ & NA48/2~\cite{NA482:2009pfe}\\
$\mu\mu$ & $-0.575\pm 0.039$ & $-0.813\pm 0.145$ & NA48/2~\cite{NA482:2010zrc}\\
\end{tabular}
\quad 
\begin{tabular}{cccr}
\multicolumn{4}{c}{\emph{Current situation}}\\\hline\hline
Channel & $a_+$ & $b_+$ & Reference \\ \hline
\multirow{2}{*}{$ee$} & \multirow{2}{*}{$-0.561\pm 0.009$} & \multirow{2}{*}{$-0.694\pm 0.040$} & \multirow{2}{*}{comb.~\cite{DAmbrosio:2018ytt}}\\
 &  &  & \\
$\mu\mu$ & $-0.592\pm 0.015$ & $-0.699\pm 0.058$ & NA62~\cite{Bician:2020ukv}\\
\end{tabular} 
}
\caption{\small Summary of the estimation of vector form factors for $K^+\to \pi^+ \ell \bar\ell$. The left panel gives the historical progression and the right panel gives the current status.}
\label{tab:a+b+}
\end{table}

In the past, the extraction of $a_+$ for the electron and the muon has been done from  experimental data in refs.~\cite{E865:1999ker,NA482:2009pfe,NA482:2010zrc} as shown  in the left panel of table~\ref{tab:a+b+}. The central values and the corresponding uncertainties led to the conclusion of the measurements being consistent with lepton flavour universality conservation. 

\begin{figure}[t]
\begin{center}
\includegraphics[width=0.48\textwidth]{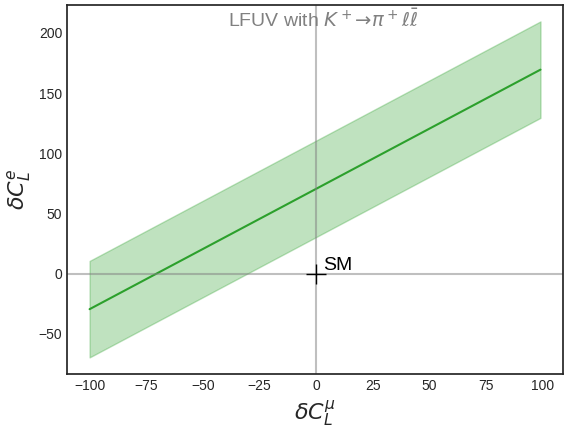}
\caption{\small
Region consistent with the estimation of the LFUV variable in $K^+ \to \pi^+ \ell \bar{\ell}$ decays.
\label{fig:K_LFUV}}
\end{center}
\end{figure}

The most recent determination of the vector form factor for muons is from the  NA62 experiment~\cite{Bician:2020ukv} as given in the right panel in table~\ref{tab:a+b+}. Comparing with the number due to NA48/2~\cite{NA482:2010zrc}, we find that while the central value remains largely unchanged, the uncertainties have been reduced by more than a factor of 2. With the ongoing program, further improvements are expected in the future. 
For the electron sector, there are two measurements by the E865~\cite{E865:1999ker} and NA48/2~\cite{NA482:2009pfe} experiments. 
The  parameters of the form factor~$V_+(z)$ are individually fitted to the two available data sets.
The data sets are in agreement for most values of $z$ except for those around $z=0.3$ \cite{DAmbrosio:2018ytt}. However, a rescaling of the errors in that region by a factor of about 2.5 leads to an agreement between the two. Thus the combination, using the rescaling at $z=0.3$ lead to the numbers in the right column of  table~\ref{tab:a+b+}. Similar to figure~\ref{fig:Kppinunu_CLeCLmu}, we represent the results in $(\delta C_e,\delta C_\mu)$ plane in figure~\ref{fig:K_LFUV}. Using the updated values in table~\ref{tab:a+b+}  and eq.~\ref{eq:K_LFUV}, we obtain  the region   consistent with the measurements. The SM point $(0,0)$ is about $1.5\sigma$ away from the region consistent with the measured values.
 As illustrated by the  green band (within $1\sigma$ for one degree of freedom), the non-universality can be explained by a broad range of values. However, a key point to note is  the requirement of a zero electron contribution suggests an unreasonably large contribution from the Wilson coefficient for the muon.
 
\subsection{BR(\texorpdfstring{$K_{S,L}\to \mu  \bar{\mu}$}{KS,KL -> mu mu}), their interference and theoretical errors}
\label{sec:Ksmumu}
The branching ratio of the $K_{S,L} \to \mu  \bar{\mu}$ decays are interesting in different aspects. The precise  determination of $K_L\to\mu \bar{\mu}$ \cite{PDG2020} in addition to the 
ongoing efforts in $K_S\to\mu \bar{\mu}$ by LHCb \cite{LHCb:2020ycd} prompt the inclusion of these decay modes in the observables of interest. 
The analytic form of the branching fractions, in the absence of right-handed and (pseudo)scalar operators, and suited to our notation is given 
by~\cite{Isidori:2003ts,Chobanova:2017rkj} 
\begin{align}
\label{eq:brKSmumu}
{\rm BR}(K_{S} \to \mu  \bar{\mu} ) &=  \tau_{S}  \frac{ f_K^2 m_K^3 \beta_{\mu}} { 16 \pi} 
 \left\{ \beta_\mu^2 \left|N_{S}^{\rm LD}\right|^2  + \left( \frac{2m_\mu}{m_K}\frac{G_F \alpha_e }{\sqrt{2}\pi}  \right)^2  {\rm Im}^2
\left[ -\lambda_c \frac{Y_c}{s_W^2} +\lambda_t C_{10}^{\ell}\right]   \right\},
\end{align}
and for the branching ratio of the $K_{L} \to \mu  \bar{\mu}$ decay we have
\begin{align}
\label{eq:brKLmumu}
{\rm BR}(K_{L} \to \mu  \bar{\mu} ) &=  \tau_{L}  \frac{ f_K^2 m_K^3 \beta_{\mu}} { 16 \pi} 
 \left| N_{L}^{\rm LD} - \left( \frac{2m_\mu}{m_K}\frac{G_F \alpha_e }{\sqrt{2}\pi}  \right)
{\rm Re}\! \left[ -\lambda_c \frac{Y_c}{s_W^2} +\lambda_t C_{10}^{\ell} \right] \right|^2,
\end{align}
where the short-distance SM contribution is given by $Y_c$
and $C_{10,{\rm SM}}^{\ell}=C_{10,{\rm SM}} ={-Y(x_t)}/{ s_W^2}$ (see appendix~\ref{app:Yxt}) and
the long-distance contributions as extracted in~\cite{Chobanova:2017rkj} from~\cite{Ecker:1991ru, Isidori:2003ts, DAmbrosio:2017klp, Mescia:2006jd}~(see also~\cite{Quigg:1968zz,Martin:1970ai,Savage:1992ac,DAmbrosio:1992zqm,DAmbrosio:1996lam,Valencia:1997xe,DAmbrosio:1997eof,Cirigliano:2011ny}):  
\begin{align}
N_{S}^{\rm LD} & =  (-2.65 + 1.14 i )\times 10^{-11} \textrm{\,(GeV})^{-2}\,,\\
N_{L}^{\rm LD} & = \pm   \left[0.54(77) - 3.95 i\right]\times 10^{-11} \textrm{\,(GeV})^{-2}\,,
\label{eq:LDKmumu}
\end{align}
with $N_{L}^{\rm LD}$ having an unknown sign (see appendix~\ref{app:NLLD} for further details).
Our SM evaluation for $K_{S,L}\to\mu \bar{\mu}$ using the updated inputs with  the corresponding uncertainties  are given below:
\begin{align}
\text{BR}(K_S\to \mu \bar{\mu})^{\rm SM}       &= (5.15\pm1.50)\times 10^{-12}\,,    \\[4pt]
\text{BR}(K_L \to \mu  \bar{\mu})^{\rm SM} &=  
\begin{cases}
 {\rm LD}(+)\!:\; \left(6.82^{+0.77}_{-0.24}\pm0.04\right)\times 10^{-9}\,,\\[4pt]
 {\rm LD}(-)\!:\; \left(8.04^{+1.46}_{-0.97}\pm0.09\right)\times 10^{-9}\,. 
\end{cases} 
\end{align}

The estimation for BR($K_S\to \mu \bar{\mu}$) is
in perfect agreement with  past literature
\cite{Ecker:1991ru,Isidori:2003ts,Gorbahn:2006bm,DAmbrosio:2017klp}. The corresponding evaluation of $K_L\to\mu \bar{\mu}$ leads to `two' SM predictions, each corresponding to a given sign of  $N_{L}^{\rm LD}$.
The point of interest is in the evaluation of the corresponding asymmetric error emerging mainly from the uncertainty in the long-distance contribution. The existing computations quote symmetric errors  which leads to a 1$\sigma$ agreement for both signs of $N_{L}^{\rm LD}$ with the corresponding experimental measurement. In this work, we investigate the asymmetric  uncertainty  of the branching fraction. 
\begin{figure}[t!]
\begin{center}
\includegraphics[width=0.55\textwidth]{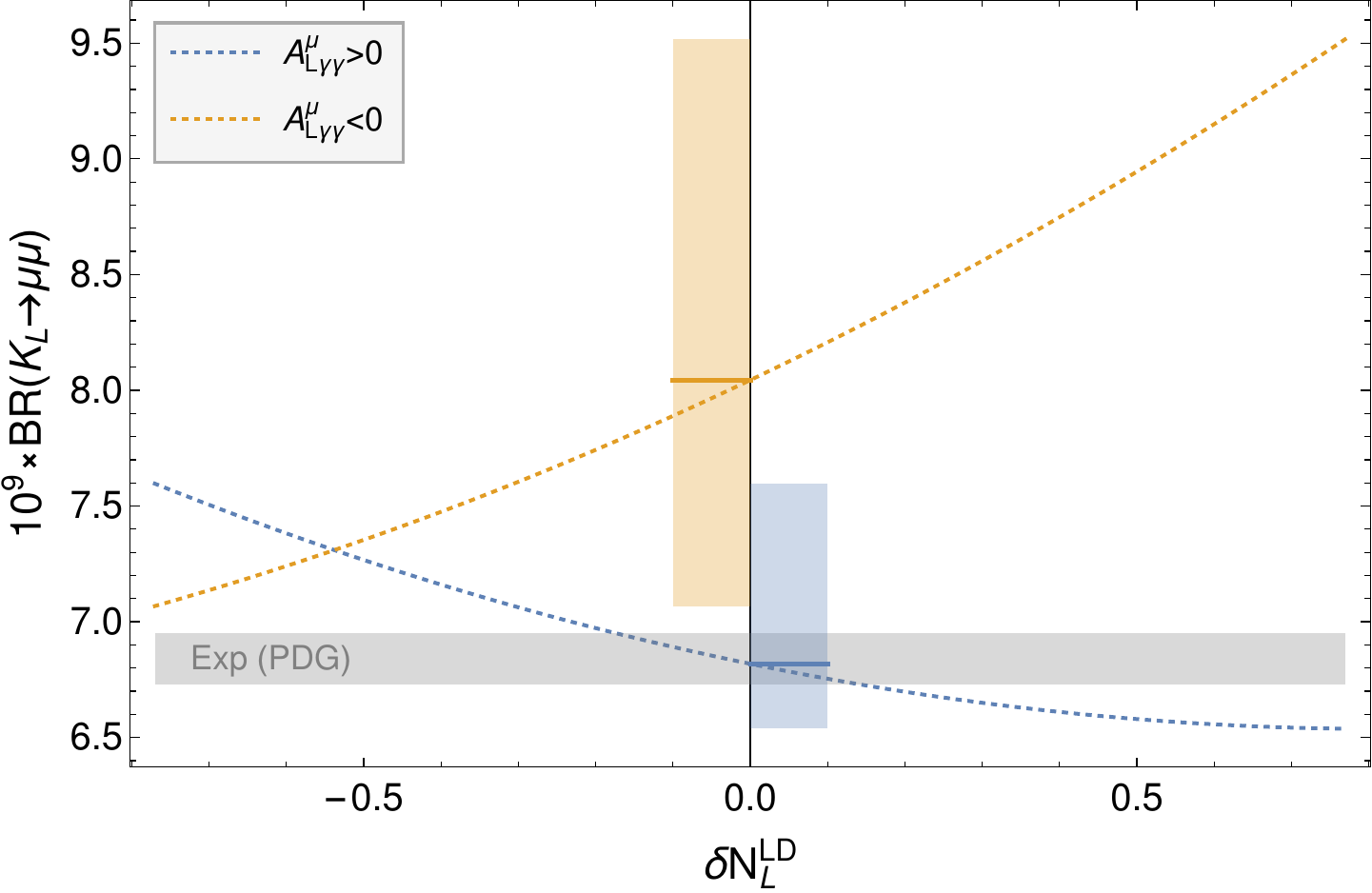}
\caption{\small 
The SM theoretical error of BR($K_L \to \mu  \bar{\mu}$) due to the uncertainty in the long-distance contribution. \label{fig:KLmumu_err}}
\end{center}
\end{figure}
The dotted lines in figure~\ref{fig:KLmumu_err} show the variation in $K_L\to\mu \bar{\mu}$ as $N_{L}^{\rm LD}$ is varied over the 1$\sigma$ interval: blue (orange) corresponds to the +($-$) sign of $N_{L}^{\rm LD}$. 
The SM central values are represented by orange and blue horizontal lines in the coloured regions.
The grey shaded region is the experimental measurement within the allowed $1\sigma$ error bars.
Considering all inputs and assuming a Gaussian distribution of the errors of the inputs, we estimate the errors for each sign of $N_{L}^{\rm LD}$ with a Monte Carlo analysis (see appendix~\ref{app:error} for details). 
There are some points that stand out at this juncture:~A)~asymmetric pattern of the errors about the central values and~B)~minor disagreement (slightly above $1\sigma$) of the negative sign of $N_{L}^{\rm LD}$ with the experimental measurement.

\begin{figure}[t!]
\begin{center}
\includegraphics[width=0.48\textwidth]{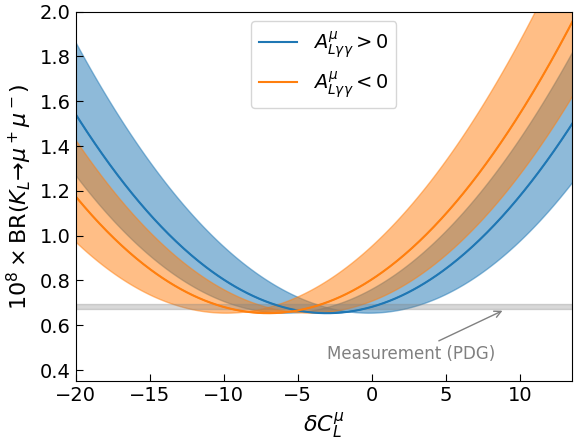}\quad
\includegraphics[width=0.48\textwidth]{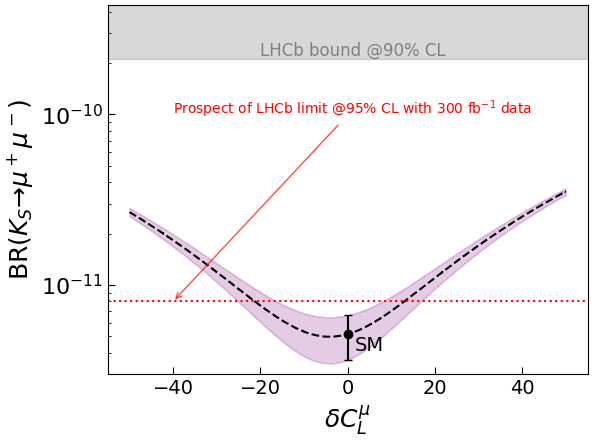}
\caption{\small
BR($K_L\to \mu\bar{\mu}$) as a function of $\delta C_L^\mu(\equiv \delta C_9^\mu=-\delta C_{10}^\mu)$ assuming both possible signs for the long-distance contribution from $A_{L\gamma\gamma}^\mu$ on the left panel. BR($K_S\to \mu \bar{\mu}$) as a function of NP contributions in $\delta C_L^\mu$ on the right panel. In the left (right) panel, the grey band indicates  the experimental measurement (upper limit) while the coloured bands correspond to the theoretical uncertainties. The LHCb bound and prospect for BR($K_S\to \mu \bar{\mu}$) are from ref.~\cite{LHCb:2020ycd} and ref.~\cite{LHCb:2018roe}, respectively.
\label{fig:KLmumu_CL}}
\end{center}
\end{figure}

The large uncertainty in the long-distance contribution results in quite asymmetric uncertainties in the branching ratio of $K_L \to \mu  \bar{\mu}$ (see figure~\ref{fig:KLmumu_err}). 
The asymmetry is also reflected in the computation of $K_L\to\mu \bar{\mu}$ with the inclusion of NP. The left plot of figure~\ref{fig:KLmumu_CL} gives the computation of BR$(K_L\to\mu \bar{\mu})$ as a function of $\delta C_L^\mu (\equiv -\delta C_{10}^\mu)$ for both signs of the long-distance contributions. The widths of the coloured bands represent the $1\sigma$ theoretical uncertainties. The band has a non-uniform width which appears to be pinched at $\delta C_L\simeq -5$ corresponding to the negligible lower uncertainty at that point.
As noted before and in table~\ref{tab:data}, the experimental measurement of BR($K_{L} \to \mu \bar{\mu}$) is precise with less than 2\% uncertainty and is shown by the grey band in the figure.
Thus, irrespective of the large theory uncertainty and the unknown sign of the long-distance contributions from $A_{L\gamma\gamma}^\mu$, NP contribution to $\delta C_L^\mu$ is limited to the $[-13.4,3.4]$ range at 1$\sigma$.

The measurement of  BR($K_{S} \to \mu \bar{\mu}$) is in its preliminary stages. This is illustrated by the right plot of figure~\ref{fig:KLmumu_CL}, where the grey band gives the current upper bound from LHCb \cite{LHCb:2020ycd} and the pink region gives the computation for a broad range of values of $\delta C_L$. The varying width corresponds to the varying uncertainty as a function of $\delta C_L$.
Note that even with the projected reach of LHCb with an integrated luminosity of 300 fb$^{-1}$ of data, this decay mode on its own is not sensitive to the regions of $\delta C_L$ permitted by BR($K_{L} \to \mu \bar{\mu}$). This prompts us to include the interference effects with $K_{L} \to \mu \bar{\mu}$ in $K_S\to\mu \bar{\mu}$ which was proposed in~\cite{DAmbrosio:2017klp} (see also~\cite{Dery:2021mct}). 
However, it should be noted that the future measurement of BR($K_S \to \mu \bar\mu$) at LHCb will be a powerful probe of New Physics scenarios involving scalar and pseudoscalar contributions~\cite{Chobanova:2017rkj}.

\begin{figure}[t!]
\begin{center}
\includegraphics[width=0.48\textwidth]{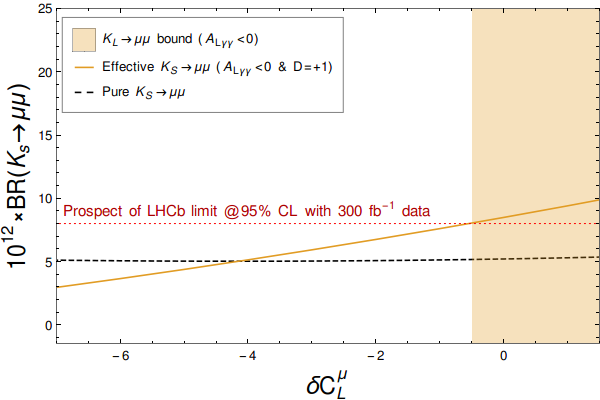}\quad\includegraphics[width=0.48\textwidth]{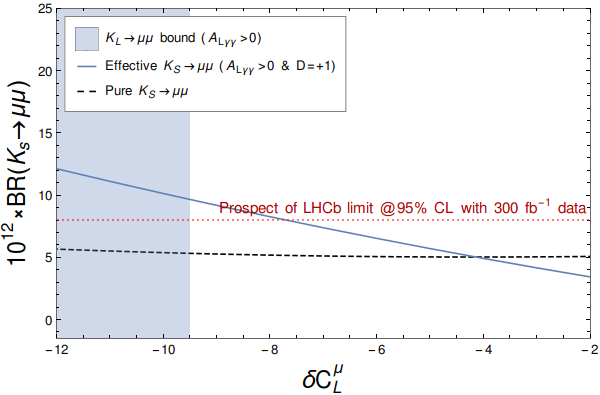}\\
\includegraphics[width=0.48\textwidth]{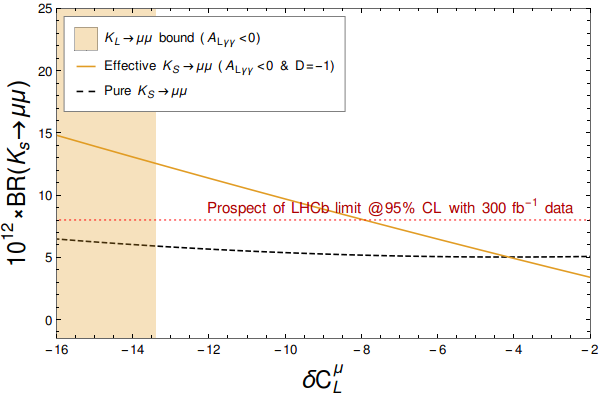}\quad
\includegraphics[width=0.48\textwidth]{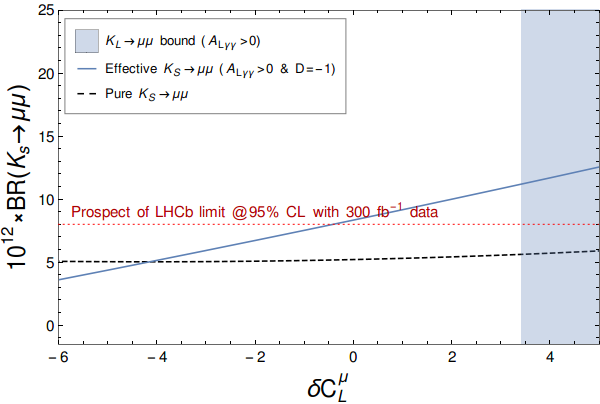}
\caption{\small 
The impact of the interference of $K_L\to \mu\bar{\mu}$ on the effective branching fraction of $K_S\to\mu\bar{\mu}$. The left and the right panels correspond to negative and positive signs for $A_{L\gamma\gamma}$, respectively. The upper and lower panels correspond to dilution factor $D=+1$ and $D=-1$, respectively.
\label{fig:KSmumuEff}}
\end{center}
\end{figure}

Figure~\ref{fig:KSmumuEff} gives the impact of the interference of $K_L\to \mu\bar{\mu}$ on the effective branching fraction of $K_S\to\mu\bar{\mu}$. The results are presented for two extreme values of the dilution factor $D=\tfrac{K^0-\bar K^0}{K^0+\bar K^0}$, which is a measure of the initial asymmetry of $K^0$ and $\bar K^0$. The left (right) column corresponds to $A_{L\gamma\gamma}<0$ ($A_{L\gamma\gamma}>0$) and the shaded region in each plot represents the region ruled out by the measurement of BR($K_L\to\mu \bar\mu$). Comparing with figure~\ref{fig:KLmumu_CL}, we note that the inclusion of interference effects makes the $\delta C_L^\mu\sim\mathcal{O}(1)$ region accessible to the high luminosity phase of LHCb. 

\subsection{\texorpdfstring{$K_L\to \pi^0\ell \bar{\ell}$}{KL -> pi0 ll}}
\label{sec:KLtopill}
These processes have long been considered a smoking gun for the detection of direct CP violation. The description is composed of three contributions: the CP-conserving long-distance contribution, which is through the two-photon process $K_L\to \pi^0\gamma^*\gamma^*$, an indirect CP-violating contribution proportional to the CP-conserving process BR($K_S\to\pi^0 \ell  \bar{\ell}$) and  $\epsilon$  which parameterises the $K^0-\bar K^0$ mixing and finally  the direct CP-violating contribution.  
Considering the different contributions, the branching fraction of $K_L \to \pi^0 \ell \bar{\ell}$ can be expressed as~\cite{Buchalla:2003sj,Isidori:2004rb,Mescia:2006jd}~(see also~\cite{Martin:1970ai, Ecker:1987qi, Donoghue:1987awa, Ecker:1987hd, Ecker:1987fm, Sehgal:1988ej, Flynn:1988gy, Cappiello:1988yg, Morozumi:1988vy, Ecker:1990in, Savage:1992ac, Cappiello:1992kk, Heiliger:1992uh, Cohen:1993ta, Buras:1994qa, DAmbrosio:1996kjn, Donoghue:1997rr, KTeV:1999gik, Murakami:1999wi, KTeV:2000amh, Diwan:2001sg, Gabbiani:2001zn, Gabbiani:2002bk, NA48:2002xke})
\begin{align}
 {\rm BR}(K_L \to \pi^0 \ell \bar{\ell}) = \left( C_{\rm dir}^\ell \pm C_{\rm int}^\ell|a_S| + C_{\rm mix}^\ell|a_S|^2 + C_{\gamma \gamma}^\ell  \right)\cdot 10^{-12}\,,
\end{align}
with $|a_S| = 1.20 \pm 0.20$, extracted from experimental results on the branching fractions  of $K_S \to \pi^0 e  \bar{e}$ and $K_S \to \pi^0 \mu  \bar{\mu}$.
The numerical values of the  different components are given in~\cite{Mescia:2006jd} as collected below:
\begin{table}[H]
\renewcommand{\arraystretch}{1.3}
\centering
\scalebox{0.9}{
\begin{tabular}{c|c|c|c|c}
 & $C_{\rm dir}^\ell$ & $C_{\rm int}^\ell$ & $C_{\rm mix}^\ell$ & $C_{\gamma \gamma}^\ell$ \\
\hline
$\ell=e$ & $(4.62\pm 0.24) (w_{7V}^2 + w_{7A}^2)$ &  $(11.3\pm 0.3) w_{7V}$ &  $14.5\pm 0.5$ &  $ \approx 0$ \\
$\ell=\mu$ & $(1.09\pm 0.05) (w_{7V}^2 + 2.32w_{7A}^2)$ &  $(2.63\pm 0.06) w_{7V}$ &  $3.36 \pm 0.20$ &  $5.2 \pm 1.6$ \\
\end{tabular}
}
\end{table}

\noindent
where $C_{\rm dir}$  corresponds to the direct CP-violating term determined by short-distance contributions proportional to Im($\lambda_t^{sd}$) in the SM (and minimal flavour violating scenarios~\cite{Chivukula:1987fw, Hall:1990ac, Rattazzi:2000hs, Buras:2000dm, DAmbrosio:2002vsn}). The  $C_{\rm mix}^\ell$ term  indicates the indirect CP-violating contribution and $C_{\rm int}^\ell$ corresponds to the interference between the direct and indirect CP-violating contributions. The sign of the latter contribution is unclear, although constructive interference is preferred~\cite{Buchalla:2003sj}.\footnote{In this paper, we consider constructive interference when investigating NP.}  Finally, the $C_{\gamma \gamma}^\ell$ ($\equiv C_{\rm CPC}$) term corresponds to the CP-conserving contribution from the two-photon intermediate states which can be deduced from the measurement of the $K_L\to\pi^0\gamma\gamma$ spectrum~\cite{NA48:2002xke}. The fact that this contribution is negligible for the electron mode strengthens the idea that the electron mode, in particular, could be an incontrovertible signal for the presence of direct CP violation. As indicated in \cite{Buchalla:2003sj},  40$\%$ of the contribution to the branching fraction is due to the clean short-distance physics, primarily driven by the interference with the indirect CP-violating part.

The $C^\ell_{\rm dir}$ and $C^\ell_{\rm int}$ contributions are parameterised by the 
factors $w_{7V,7A}$ which encode the short-distance SM and NP effects. They are defined as (see e.g.~\cite{Bobeth:2016llm})
\begin{align}
 w_{7V} = \frac{1}{2\pi}{\rm Im}\left[ \frac{\lambda_t^{sd}}{1.407\times 10^{-4}} C_9  \right]\,,\quad
 w_{7A} = \frac{1}{2\pi}{\rm Im}\left[ \frac{\lambda_t^{sd}}{1.407\times 10^{-4}} C_{10}  \right]\,,
\end{align}
where $1.407\times 10^{-4}$ corresponds to the input used by~\cite{Mescia:2006jd} for $\lambda_t^{sd}$. Using  the updated inputs in table~\ref{tab:inputs} we find for constructive (destructive) interference
\begin{align}
{\rm BR}^{\rm SM}(K_L \to \pi^0 e  \bar{e}) &= 3.46^{+0.92}_{-0.80} \left( 1.55^{+0.60}_{-0.48} \right) \times 10^{-11}\,,\\[8pt]
{\rm BR}^{\rm SM}(K_L \to \pi^0 \mu  \bar{\mu}) &= 1.38^{+0.27}_{-0.25} \left( 0.94^{+0.21}_{-0.20} \right) \times 10^{-11}\,,
\end{align}
while current experimental bounds from KTeV~\cite{KTeV:2003sls,KTEV:2000ngj} at 90\% confidence level (CL) are one order of magnitude larger 
\begin{align}
{\rm BR}^{\rm exp}(K_L \to \pi^0 e  \bar{e}) &< 28 \times 10^{-11}\qquad \text{at 90\% CL}\,,\\
{\rm BR}^{\rm exp}(K_L \to \pi^0 \mu  \bar{\mu}) &< 38 \times 10^{-11}\qquad \text{at 90\% CL}\,.
\end{align}

It is expected that in  the hybrid phase of the future CERN kaon program these decay modes are going to be observed~\cite{Goudzovski:2022vbt}. 
Currently, the dominating theoretical uncertainty is due to $|a_S|$~\cite{Buchalla:2008jp} followed by the uncertainty on the two-photon intermediate state contribution in the muon mode which for destructive interference is as large as the uncertainty due to $|a_S|$. 
The prospect of the $|a_S|$ form factor determination is 10\% statistical precision with LHCb Upgrade~II~\cite{LHCb:2018roe,Alves:2018npj}.

\begin{figure}[htbp]
\begin{center}
\includegraphics[width=0.6\textwidth]{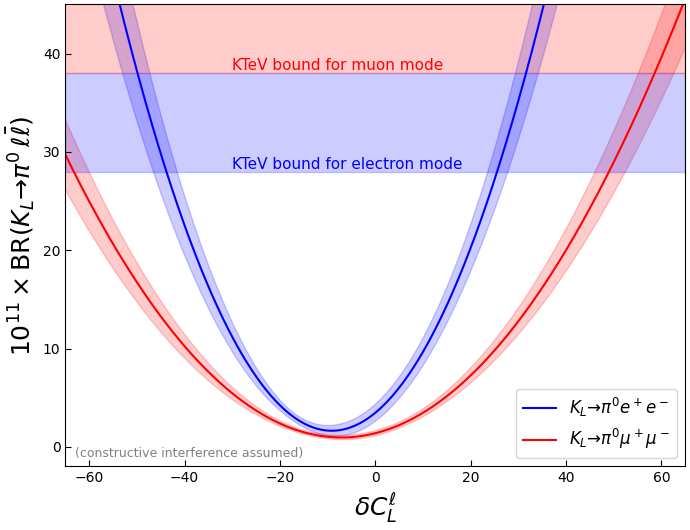}
\caption{\small
BR($K_L \to \pi^0 \ell \bar\ell$) as a function of $\delta{C_L^\ell} (\equiv C_9^\ell=-C_{10}^\ell)$ assuming constructive interference between direct and indirect CP-violating contributions.
\label{fig:KLpill_pos}}
\end{center}
\end{figure}

The effect of NP in $\delta C_L$ on the branching fraction of $K_{L} \to \pi^0 \ell \bar{\ell}$ is shown in figure~\ref{fig:KLpill_pos} for the electron and the muon sectors.
Since currently there are only upper bounds from experiments, these two observables do not put stringent constraints on possible NP contributions. For the muon sector, the upper bound gives a much weaker constrain compared to BR($K_L \to \mu \bar{\mu}$) as given in the previous subsection. Nonetheless, it is impressive that for the electron sector the current upper limit on BR($K_{L} \to \pi^0 e  \bar{e}$) which is about one order of magnitude larger than the SM prediction is already compatible with what is obtained by BR($K^+\to \pi^+ \nu \bar{\nu}$),  indicating $\delta C_L^e \lesssim 28$ at 90\% CL.

\section{Global picture}
\label{sec:global}
Table~\ref{tab:data} summarises the results of the last section. The first column gives our evaluated SM value and the second column is the current experimental precision. The last column gives the experimental projection for the measurement of these observables and is detailed in section~\ref{sec:global1}.
\begin{table}[h]
\renewcommand{\arraystretch}{1.39}
\begin{center}
\setlength\extrarowheight{1pt}
\scalebox{0.616}{
\begin{tabular}{|l|l|ll|l|}\hline
\bf{Observable} & \bf{SM prediction}& \bf{Exp results} & \bf{Ref.}& \bf{Experimental Err. Projections} \\ \hline
BR$(K^+\to \pi^+\nu\nu)$    & $(7.86 \pm 0.61)\times 10^{-11}$  & $(10.6^{+4.0}_{-3.5} \pm 0.9 ) \times 10^{-11}$ & \cite{NA62:2021zjw}& 10\%(@2025)\,5\%(CERN; long-term)~\cite{Goudzovski:2022vbt}  \\
BR$(K^0_L\to \pi^0\nu\nu)$  & $(2.68 \pm 0.30) \times 10^{-11}$ & $ <3.0\times 10^{-9} $ @$90\%$ CL & \cite{Ahn:2018mvc}& $20\%$(CERN; long-term ~\cite{Goudzovski:2022vbt})\, 15\% (KOTO~\cite{NA62:2020upd}) \\
LFUV($a_+^{\mu\mu}-a_+^{ee}$)&\multicolumn{1}{c|}{0}&$-0.031\pm 0.017$&\cite{DAmbrosio:2018ytt,Bician:2020ukv}&$\pm0.007$ (assuming $\pm0.005$ for each mode)\\
BR$(K_L\to \mu\mu)$ ($+$)   & $(6.82^{+0.77}_{-0.29})\times 10^{-9}$    & \multirow{2}{*}{$(6.84\pm0.11)\times 10^{-9}$} & \multirow{2}{*}{\cite{PDG2020}} &  \multirow{2}{*}{
{\small experimental uncertainty kept to current value}}\\
BR$(K_L\to \mu\mu)$ ($-$)   &  $ (8.04^{+1.47}_{-0.98})\times 10^{-9}$    &  & & 
\\
BR$(K_S\to \mu\mu)$         & $(5.15\pm1.50)\times 10^{-12}$    & $ < 2.1(2.4)\times 10^{-10}$ @$90(95)\%$ CL & \cite{LHCb:2020ycd} & $<8\times10^{-12}$ @$95\%$ CL (CERN; long-term~\cite{LHCb:2018roe})\\
BR$(K_L\to \pi^0 ee)(+)$         & $(3.46^{+0.92}_{-0.80})\times 10^{-11}$    & \multirow{2}{*}{$ < 28\times 10^{-11}$ @$90\%$ CL} & \multirow{2}{*}{\cite{KTeV:2003sls}}&  \multirow{4}{*}{observation (CERN; long-term~\cite{Goudzovski:2022vbt})}\\
BR$(K_L\to \pi^0 ee)(-)$         & $(1.55^{+0.60}_{-0.48})\times 10^{-11}$    &  &  & \\
BR$(K_L\to \pi^0 \mu\mu)(+)$         & $(1.38^{+0.27}_{-0.25})\times 10^{-11}$    & \multirow{2}{*}{$ < 38\times 10^{-11}$ @$90\%$ CL} & \multirow{2}{*}{\cite{KTEV:2000ngj}} &  {\footnotesize (we assume 100\% error)} \\
BR$(K_L\to \pi^0 \mu\mu)(-)$         & $(0.94^{+0.21}_{-0.20})\times 10^{-11}$    &  &  &  \\
\hline
\end{tabular}}
\vspace{-0.2cm}
\caption{\small 
The SM predictions and experimental status for the different observables. The  SM values and the corresponding uncertainties in the first column are evaluated using the updated inputs using {\tt{SuperIso}}~\cite{Mahmoudi:2008tp}.
In the second column, the current experimental status is given.
The third column gives the  projections in experimental sensitivity for the corresponding experiments (see also~\cite{Goudzovski:2022}).
For the case with  more than one projection, the CERN long-term ones are used.
\label{tab:data}}
\end{center}
\end{table}
It is convenient to present  a unified picture of the topics discussed thus far.
\begin{figure}[t]
\begin{center}
\includegraphics[width=0.48\textwidth]{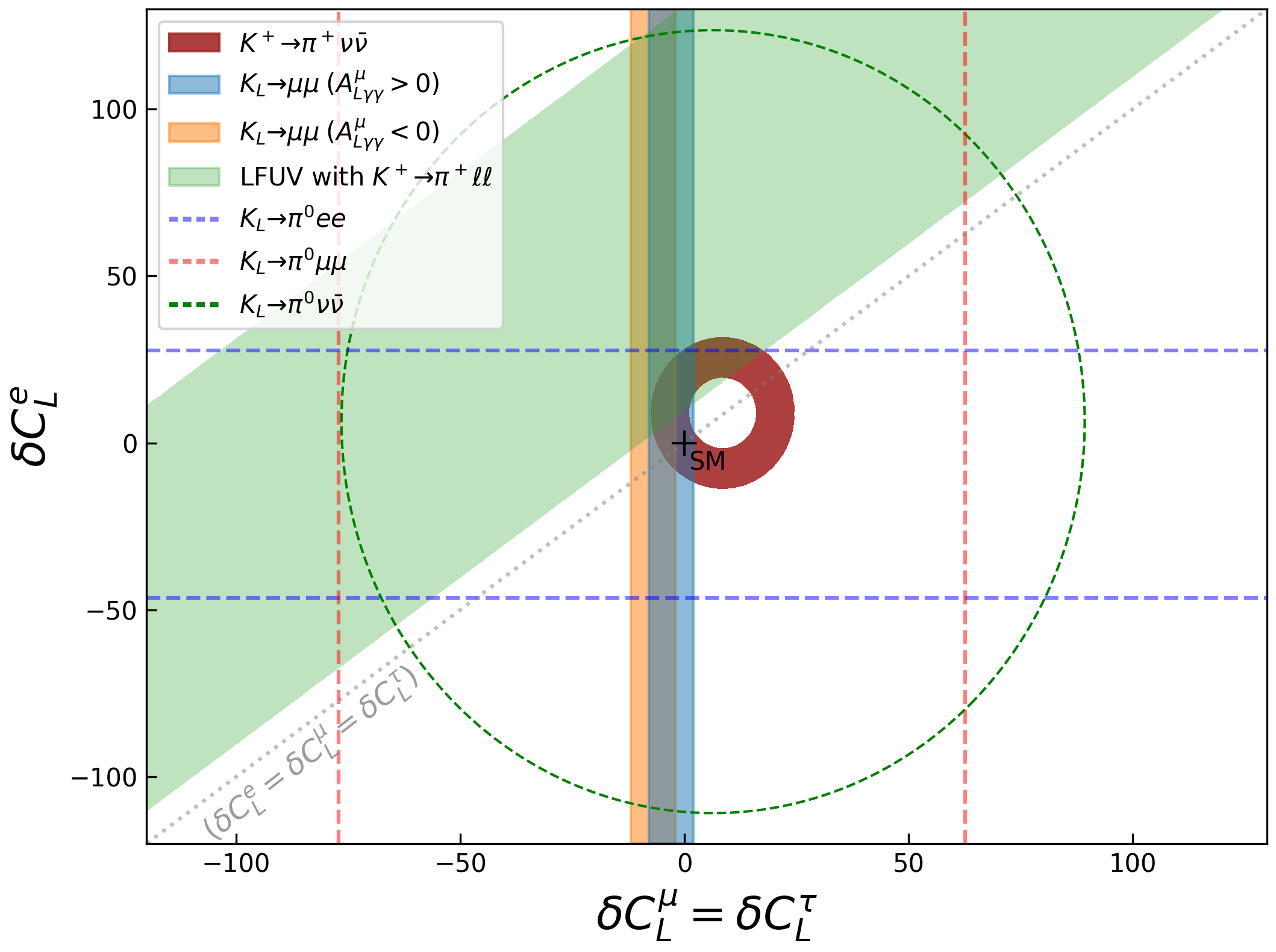}
\includegraphics[width=0.48\textwidth]{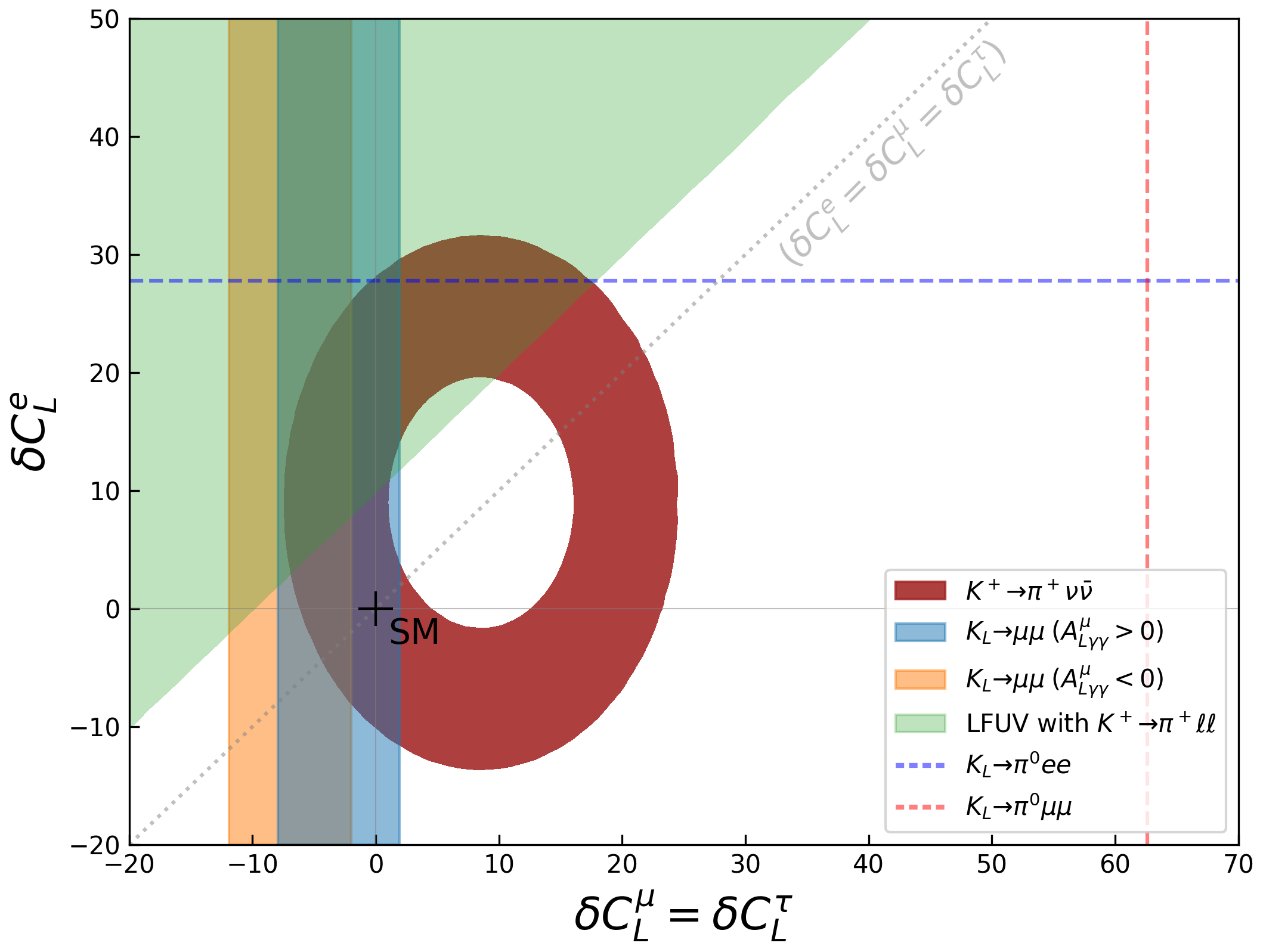}
\caption{\small
The bounds from individual observables. The right panel is the zoomed version of the left panel. The coloured regions correspond to  $68\%$ CL when there is a measurement and the dashed ones to upper limits at 90\% CL.
$K_L \to \mu \bar{\mu}$ has been shown for both signs of the long-distance contribution. For $K_L \to \pi^0 e \bar{e}$ and $K_L \to \pi^0 \mu \bar{\mu}$, constructive interference
between  direct  and  indirect  CP-violating  contributions  has  been  assumed. 
\label{fig:all_obs_individually}}
\end{center}
\end{figure}

In figure~\ref{fig:all_obs_individually}, for each individual observable, we show the $68\%$ CL regions in the $(\delta C_L^e, \delta C_L^\mu)$ plane for those observables which have been measured, as well as the 90\% CL  upper limits for the observables where there are only upper bounds.
Note the mild tension in the upper part of the 68\% CL region of $K^+\to \pi^+\nu\bar\nu$ (in maroon) with the upper bound on $K_L\to \pi^0 e \bar{e}$ (dashed blue line). This is specific to the case where we choose $\delta C_L^\mu=\delta C_L^\tau$. A contrasting picture corresponding to $ \delta C_L^e = \delta C_L^\tau$ is presented in figure~\ref{fig:all_obs_individuallyb} in appendix~\ref{app:otherpossibility}.

A zoomed version is shown in the right plot. Other decays include $K_L\to\mu \bar{\mu}$ shown with the orange (blue) band  for negative (positive) long-distance contributions. The upper bound from $K_L\to\pi^0\nu\bar\nu$ is indicated by the dashed green line. The region of intersection by the horizontal blue dashed and the vertical red dashed lines represent the parameter space allowed by $K_L\to\pi^0\ell \bar{\ell}$. While this is instructive, it would be  useful to find the region in the $(\delta C_L^\mu=\delta C_L^\tau,\,\delta C_L^e)$ plane which is consistent with all observables.
This prompts the implementation of a global fit, similar to those employed for $B$~decays. However, given the limited experimental data for most observables, we adapt a multi-prong strategy for the fits taking into account both the current and future possibilities for many of these observables.

\subsection{Global fits}
\label{sec:global1}

We begin with the definition of the $\chi^2$ statistic as follows:
\begin{equation}
 \chi^2 =  \sum_{i,j =1}^{N} \left( O_i^{\rm th}(\delta C_L^{e,\mu}) - O_i^{\rm exp} \right)
 \; C_{ij}^{-1} \; \left( O_j^{\rm th}(\delta C_L^{e,\mu}) - O_j^{\rm exp} \right)\,,
\label{eq:chisq}
\end{equation}
where $C_{i,j} $ denotes the total (theoretical and experimental) covariance matrix.
Note that an observable ${O}_i^{\rm th} $ in eq.~\ref{eq:chisq} is expressed as a function of $\delta C_L^{e,\mu}$, and  the contribution due to $\delta C_L^\tau$ 
is in principle not fixed in a two-dimensional fit to $\delta C_L^{e,\mu}$.
Henceforth, unless otherwise stated we stick to the convention $\delta C_L^{\tau}=\delta C_L^{\mu}$. 
While this choice is motivated by the convention followed in the paper thus far, the future phase  of data accumulation for each of these experiments would enable us to make a more adequate choice. To ensure clarity, we divide the discussion that follows into two parts: fits with current data and projected fits.

\subsection{Fits with current data}
\label{sec:global2}

\begin{figure}[t!]
\begin{center}
\includegraphics[width=0.48\textwidth]{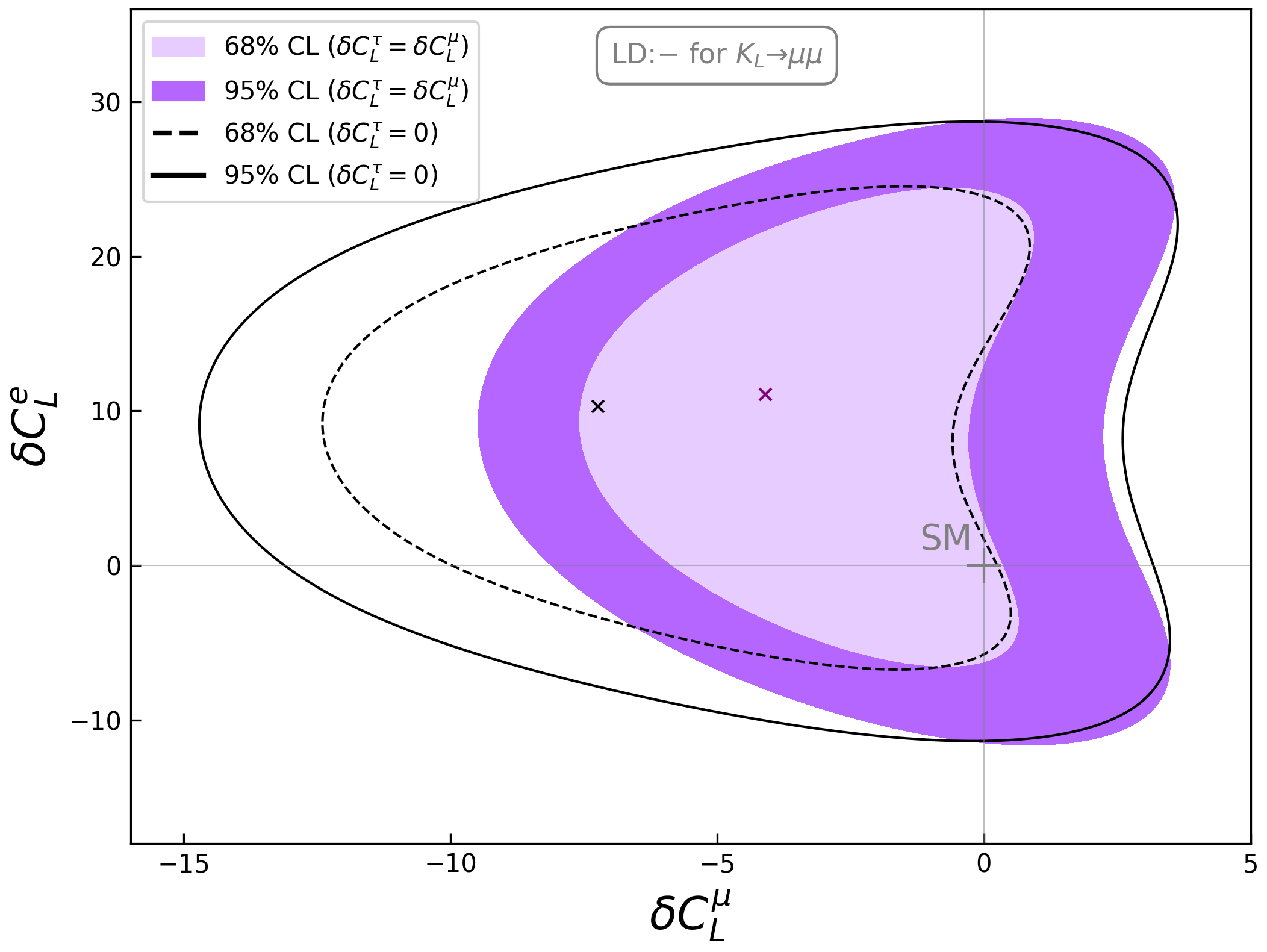}\quad
\includegraphics[width=0.48\textwidth]{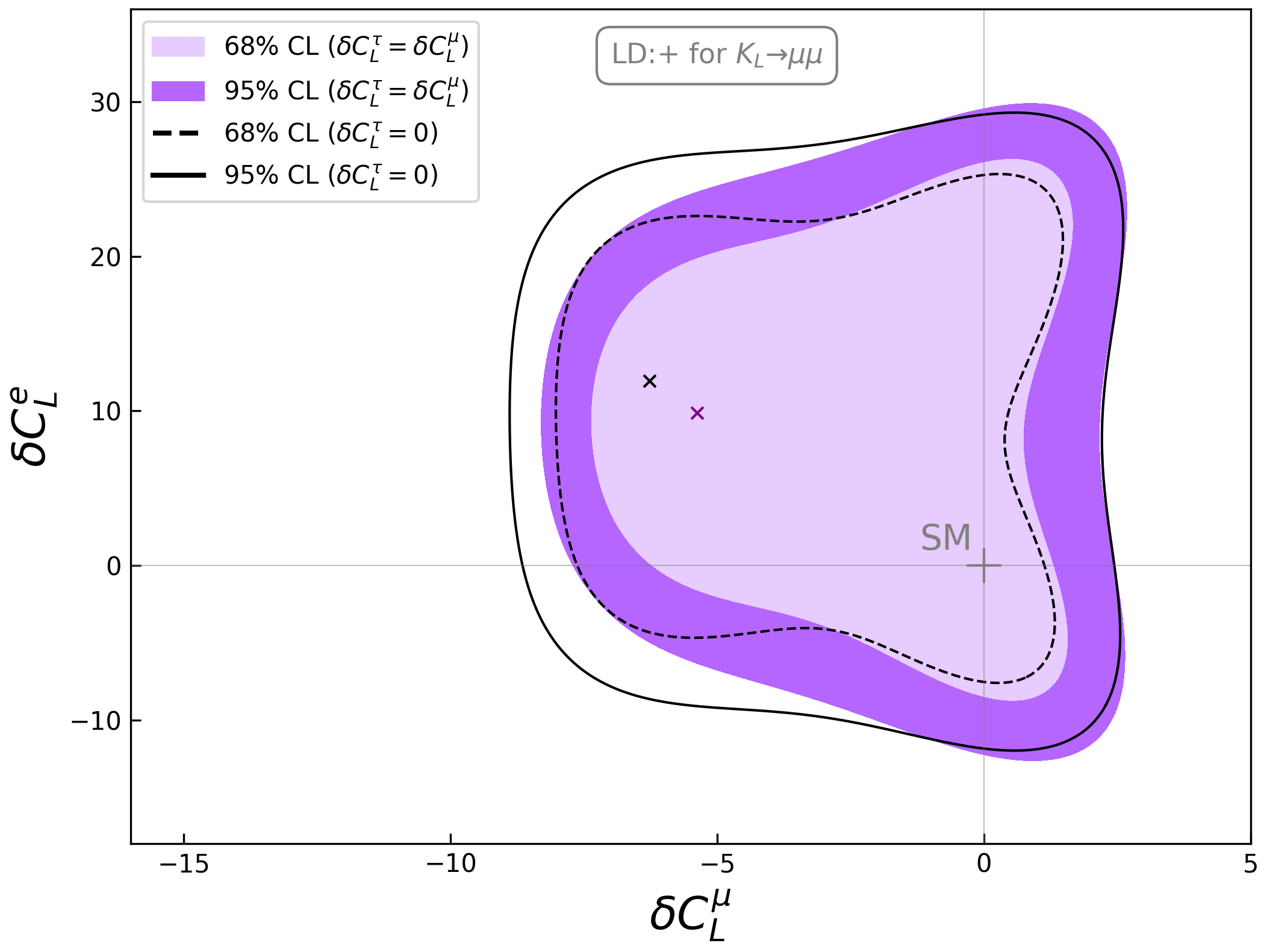}
\caption{\small
Comparison of the global fits between scenarios $\delta C_L^{\tau}=\delta C_L^{\mu}$ (purple regions) and $\delta C_L^{\tau}=~\!\!0$ (black solid and dashed lines). The fits are implemented using existing data with long-distance contribution to $K_L\to \mu \bar{\mu}$ having a negative (positive) sign, on the left (right).
\label{fig:fitcomparion}}
\end{center}
\end{figure}

We first perform a New Physics fit of $\delta C_L^e$ and $\delta C_L^\mu = \delta C_L^\tau$ to the current experimental data on rare kaon decays (collected in the second column of table~\ref{tab:data}).
The results of the fit are given in figure~\ref{fig:fitcomparion} where the 68~and~95\% CL regions are given in light and dark purple, respectively. Due to the ambiguity in the sign of the long-distance contribution from $A_{L\gamma\gamma}$ to  $K_L\to \mu \bar{\mu}$, the fit results are given for both signs with $A_{L\gamma\gamma}<0\; (>0)$ on the left (right).  
The purple cross in each plot represents the corresponding best-fit point.
While the fits are qualitatively similar, we note the appearance of a wall-shaped feature on the left side of the fit for $A_{L\gamma\gamma}>0$~(right), which is better defined than the one corresponding to $A_{L\gamma\gamma}<0$~(left). Its origin can  be traced back to the blue band in figure~\ref{fig:all_obs_individually}. But in general, the difference in shapes for the fits between positive and negative values of $A_{L\gamma\gamma}$  show that a future improvement in the sensitivity can lead to a resolution of the sign ambiguity.

One of the defining features of kaon decays is that it allows a ``relatively clean" possibility for  identifying the extent of contributions due to effective operators involving tau leptons. Note that the ``relatively clean" refers to both the status of the SM computation as well as the future projections for the measurement of observables where  operators involving the tau leptons play a direct role. These operators contribute to the computation of $K^+\to\pi^+\nu\bar \nu$ and $K_L\to\pi^0\nu\bar \nu$.
As explained in detail in section~\ref{sec:2}, both are characterised by highly precise SM computations, owing to negligible long-distance uncertainties. 
Furthermore, there exists a well-defined strategy for a highly precise measurement of both decays~\cite{Goudzovski:2022,NA62:2021zjw,Ahn:2018mvc}. It is natural to expect insight into the extent of the contributions due to the operators involving tau leptons by means of  global analyses which include these decay modes. 
This can be illustrated by a comparison of the fits
in the absence of $\delta C_L^\tau$ as shown with dashed and solid black lines corresponding to the
 68 and 95\% CL regions, respectively. 
The black cross  in figure~\ref{fig:fitcomparion} 
indicates the best-fit point for the case of  $\delta C_L^\tau = 0$.
A feature common to both signs of $A_{L\gamma\gamma}$ is that the  $\delta C_L^{\tau}=\delta C_L^{\mu}$ case prefers a smaller parameter space,  translating into a strong lower bound on $\delta C_L^\mu$. The larger spread in $\delta C_L^\mu$ for the fit with $\delta C_L^{\tau}=0$ can be attributed to the larger contribution required from the same to be consistent with the observation of $K^+\to\pi^+\nu \bar{\nu}$. Other noticeable feature is the shapes of the 68$\%$ CL contours on the right for both  plots. The depression in the centre can be attributed to the region allowed  by $K^+\to\pi^+\nu \bar{\nu}$ shown in maroon in figure~\ref{fig:all_obs_individually}. Furthermore, the wall-like feature on the left side of the global fit on the right plot reflects the  region allowed by $K_L\to\mu \bar{\mu}$ and shown by the blue band in figure~\ref{fig:all_obs_individually}. 

Visually, the left plot reveals a greater level of discrimination between the two scenarios. This is because of a relatively stronger lower bound on $\delta C^{\mu}_L$ for the negative sign of $A_{L\gamma\gamma}$ as shown by the orange region in figure~\ref{fig:all_obs_individually}. However, for either plot, the two scenarios are  statistically equivalent with the present data. A concrete discrimination may be possible with  future runs of many of these experiments. This information would be available by $\sim 2035$, at the projected end of data accumulation for NA62/KOTO thereby reaching the required precision.
This strengthens the possibility of kaon experiments offering a handle on effective operators involving tau.

\subsection{Projection fits}
\label{sec:global3}

The results of the above fit arouse a natural curiosity about the possible impact of the future runs of the experiments.
Given the preliminary stage of the run of most of these experiments, any projection on the fits requires both the possible measured value as well as the experimental precision. The latter is rather straightforward and we can adapt the intended long-term experimental precision that is true for some of the decay modes. In the case of $K^+(K_L)\to\pi^+(\pi^0)\nu\bar\nu$, there is a well-defined sensitivity goal due to NA62/KOTO. In the case of LFUV in $K^+ \to \pi^+ \ell  \bar{\ell}$, we assume  the uncertainty on $a_+^{\ell\ell}$ to become less than half the current value to reach $0.005$ for either of the electron and muon modes. For  $K_L\to\pi^0\ell \bar{\ell}$ as mentioned in section~\ref{sec:KLtopill} they are expected to be observed in the future CERN kaon program, however,  in the absence of a well-defined projection for the uncertainty, we assume 100$\%$ error.
All the numbers are collected in the last column of table~\ref{tab:data}.
A prediction of the possible measured central value, on the other hand, is not possible, especially for those with only an existing upper bound. In light of this, we present a two-faceted approach. In the first approach, labelled projection~\textbf{A}, the predicted central values for those observables with only an upper bound is projected to be the same as the SM prediction. On the other hand, the corresponding values for $K^+\to\pi^+\nu\bar\nu $, LFUV in $K^+ \to \pi^+ \ell  \bar{\ell}$ and $K_L\to\mu \bar{\mu}$ are chosen to be the same as the existing measurement.
In the second approach, labelled projection~\textbf{B}, the central values for  all of the observables are projected with the best-fit points obtained from the fits with the existing data.
 
\begin{figure}[t!]
\begin{center}
\includegraphics[width=0.48\textwidth]{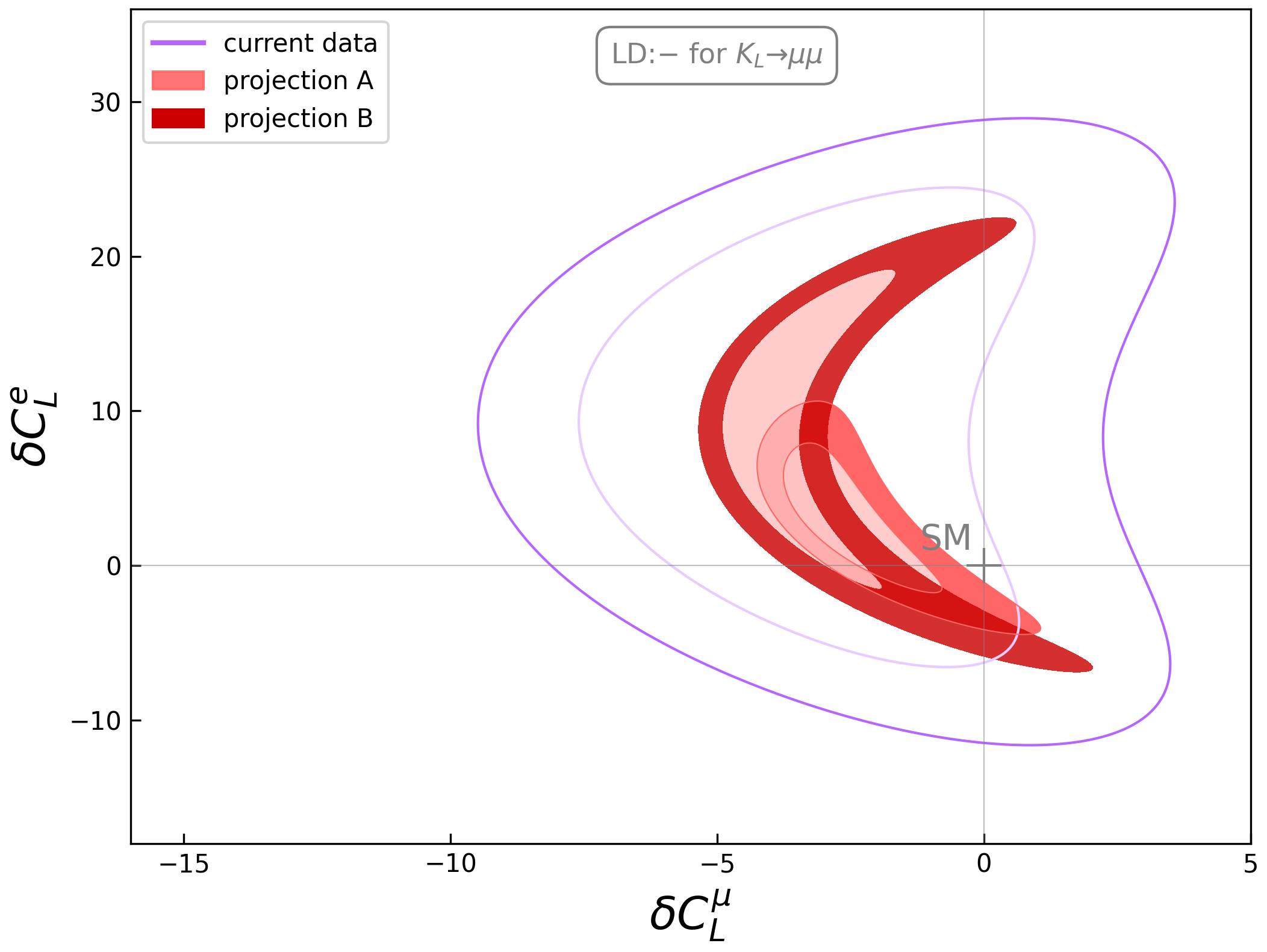}\quad
\includegraphics[width=0.48\textwidth]{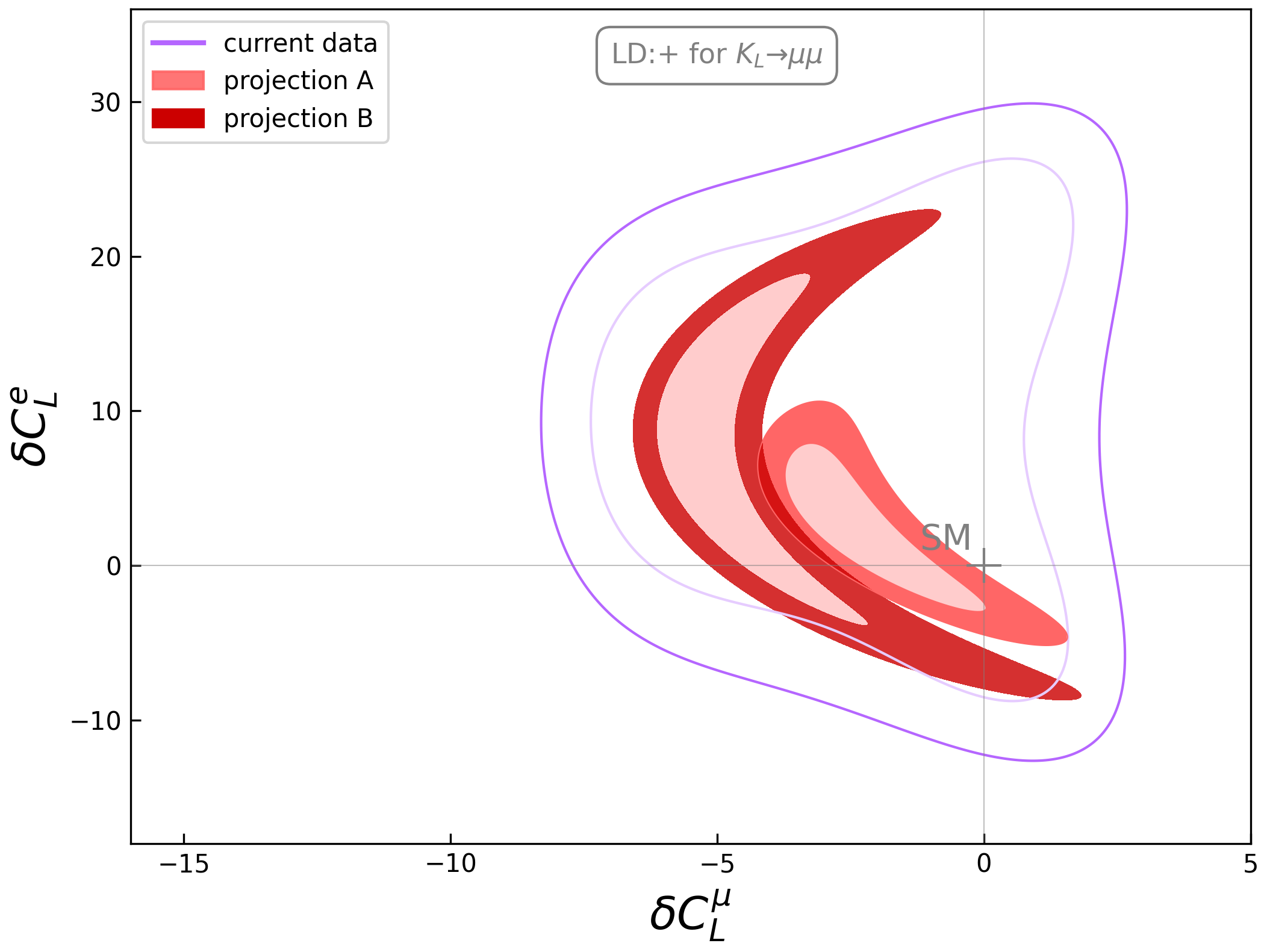}
\caption{\small
Projected fit to rare kaon observables within 68 and 95\% CL assuming negative (positive) sign for the long-distance contribution to $K_L\to \mu \bar{\mu}$ on the left (right). Lighter shades of red: projection \textbf{A} with central values from current measurements (whenever there is one) or  SM predictions (if currently not observed). Darker shades of red: projection \textbf{B}  with current best-fit point used for all projections.
The contours of the fit to current data (corresponding to the purple regions in figure~\ref{fig:fitcomparion}) are shown with purple solid lines.
\label{fig:fitprojections}}
\end{center}
\end{figure}

The results of the fits, for projections~\textbf{A} and~\textbf{B}, are illustrated in figure~\ref{fig:fitprojections} with ${A_{L\gamma\gamma}< 0\;(>)}$ on the left (right).
The 68 and 95\% CL regions are shown with lighter shades of red for projection~\textbf{A} and similarly for projection~\textbf{B} with darker shades of red. 
A~comparison with fit to current data as indicated by the purple solid contours (coinciding with the 68 and 95\% contours of the purple regions in figure~\ref{fig:fitcomparion}) reveals a significant reduction in the parameter space for both projections.
Projection~\textbf{A} leads to an overall consistency with the SM, represented by $(0,0)$, up to $3\sigma$. This can be expected with the choice of the predicted central values being the same as the SM for those observables not currently measured. Projection~\textbf{B} on the other hand predicts an overwhelming departure from the SM. This is also in line with our expectation as the best-fit points of the  fits to the current data (purple crosses in figure~\ref{fig:fitcomparion}) presented a significant departure from their corresponding SM prediction with the assumption of the  projected sensitivities.

\begin{figure}[t!]
\begin{center}
\includegraphics[width=0.48\textwidth]{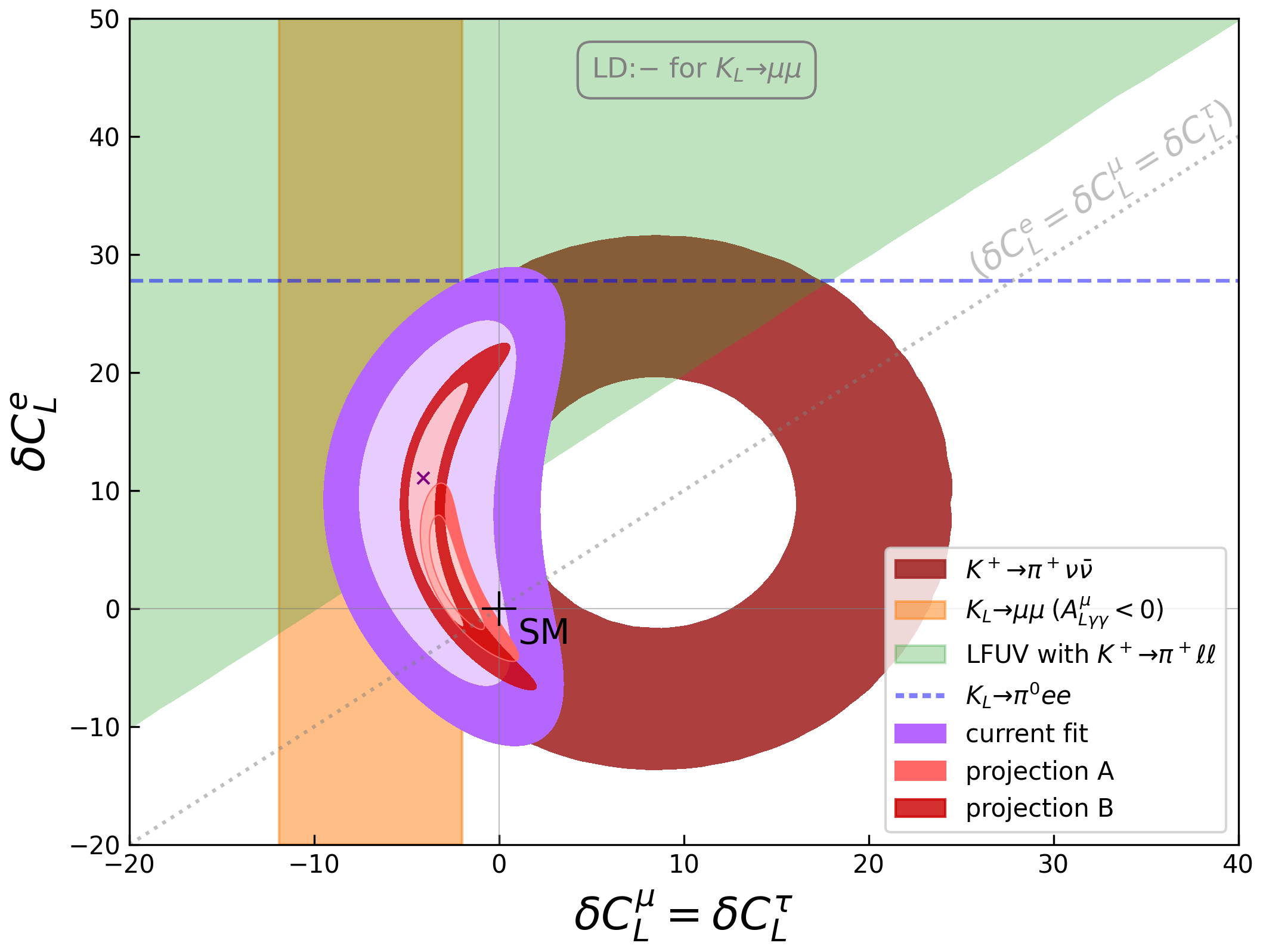}\quad
\includegraphics[width=0.48\textwidth]{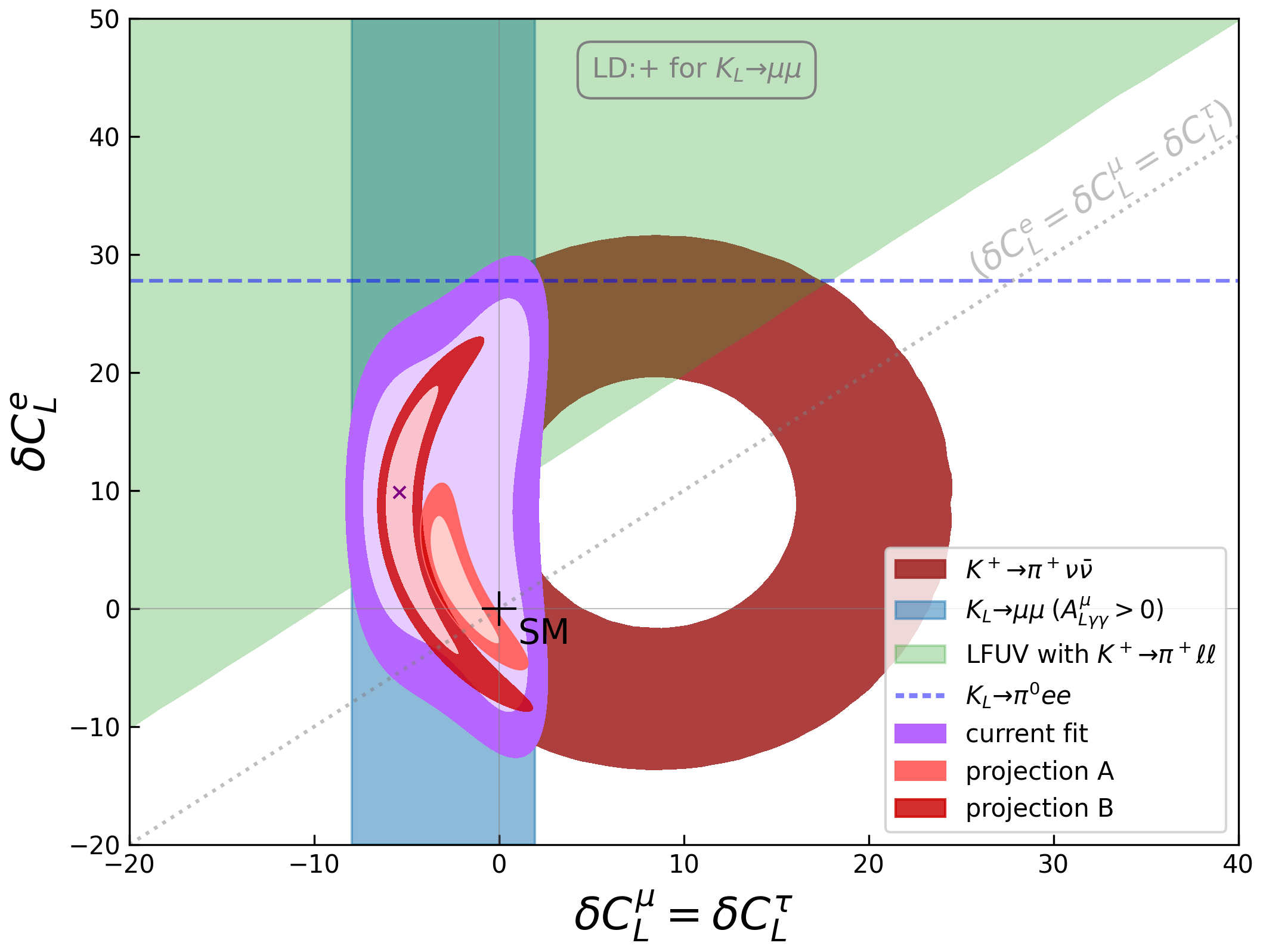}\\
\caption{\small Global fit with current data (in purple) and projection A (lighter shades of red) and projection B (darker shades of red) together with 68\% CL for individual observables with current data. The purple cross indicates the best-fit point with the current data.
\label{fig:combined}}
\end{center}
\end{figure}

The entire discussion of section~\ref{sec:global} can be conveniently encapsulated by presenting a summary plot in figure~\ref{fig:combined}. 
The left~(right) plot corresponds to $A_{L\gamma\gamma}<0~(>0)$. 
Either plot gives the results with the current fit along with the two approaches for the projected global fits: projection~A~(lighter shade of red) and B~(darker shade of red). They are overlaid on the results of figure~\ref{fig:all_obs_individually}. This gives us an illustrative understanding of the observables that are the driving force behind the fits. As expected, $K^+\to\pi^+\nu \bar{\nu}$, shown by the ``maroon-doughnut'' shape plays the most significant role in determining the shape of the regions. This is true for both the current as well as the projected fits. In an ideal case, one would expect the regions to be concentrated towards the top left for the projected fit due to the impact of the LFUV observables. However, the dominance of the theoretical errors due to $K_L\to\mu \bar{\mu}$ makes its effect to be less pronounced.

\section{Conclusions}
\label{sec:conc}
This work, motivated by $B$-anomalies, presents a possible new way to analyse  rare kaon decays looking for LFUV: we have performed global fits to the Wilson coefficients of the operators contributing to kaon decays. The road leading to the fits  is set up by a careful re-examination of the different observables involved. This includes a computation of the SM values and the theoretical uncertainties. In the case of the latter, asymmetric uncertainties were observed for some of the modes, in particular for $K_L\to\mu \bar{\mu}$. These inputs were then used to construct a global picture and develop a strategy for the fits, which were divided into two parts. The first part gave a glimpse into the existing parameter space using the current experimental information. This was then followed by a ``projection-fit" which took into account the future sensitivity and measurement goals for many of the observables while assuming realistic projections for the rest (e.g.~vector form factors in $K^+\to\pi^+\ell \bar{\ell}$). Given the uncertainty in the experimental central values for the observables for which only an upper bound exists, we adapted two methodologies:  A) assuming SM as the central values and B) assuming the best-fit values from the existing fits as the central values.
The results of the projection-fit highlighted the need to achieve a better accuracy in the theoretical computation of $K_L\to\mu \bar{\mu}$. Our $K_L\to \mu\bar\mu$ analysis leading to asymmetric errors indicates quantitatively how to strategically improve the theoretical error in order to solve the ambiguity in the sign of the long-distance contributions (can be resolved if there is an improvement by about $\sim 50\%$). 
Although not considered in our global fit,  we also demonstrated the interference of $K_S \to \mu\bar\mu$ and $K_L \to \mu\bar\mu$ besides giving a handle on the sign of the SM long-distance contributions in the latter, can be an effective probe of NP in the muon sector (in case of having an experimental setup with a large dilution factor).
This analysis presented a global picture of the physics goals that can be achieved with kaon experiments,  and the possibility to probe sensitivity to lepton flavour universality violating New Physics.

\acknowledgments

We thank Baptiste Filoche for his collaboration in the initial stages of the project. 
AMI would like to thank IP2I Lyon for the hospitality where the initial parts of the project were discussed. AMI also acknowledges support from the CEFIPRA under the  project ``Composite Models at the Interface of Theory and Phenomenology'' (Project No. 5904-C). 
GD and SN were supported in part by the
INFN research initiative Exploring New Physics (ENP).
We would like to thank Luca Lista for fruitful discussions, especially regarding the study of asymmetric uncertainties via Monte Carlo error analysis. 
We are particularly grateful to Evgueni Goudzovski and Giuseppe Ruggiero for many enlightening discussions. 
We also thank Fabio Ambrosino, Teppei Kitahara, Marc Knecht and Cristina Lazzeroni for insightful discussions.

\clearpage
\appendix

\section{Input parameters}
\label{app:inputs}
Table~\ref{tab:inputs} gives the input values used in the computation of different observables.

\begin{table}[h!]
\renewcommand{\arraystretch}{1.18}
\begin{center}
\begin{tabular}{|lr|lr|}\hline
$m_{K^\pm} = 493.677(16) $ MeV		& \cite{PDG2020} & $\lambda = 0.22650(48)$ & \cite{PDG2020}\\
$m_{K}= 497.611(13)$ MeV		& \cite{PDG2020} & $A = 0.790(17)$ & \cite{PDG2020}\\
$m_c(m_c)= 1.27(2)$ GeV		& \cite{PDG2020} & $\bar{\rho} = 0.141(17)$ & \cite{PDG2020}\\
$m_b(m_b)= 4.18 ^{+0.03}_{-0.02}$ GeV	& \cite{PDG2020} & $\bar{\eta} = 0.357(11)$ & \cite{PDG2020}  \\ 
$m_t^{pole}= 172.69(30) $ GeV		& \cite{PDG2020}  & {$\lambda_t^{sd} = \left[-3.11 (15) + i \,1.36 (7)\right]\times 10^{-4}$} &\\
$ f_K= 155.7(3)$ MeV			& \cite{FlavourLatticeAveragingGroup:2019iem} & &\\
\hline
\end{tabular}
\caption{Input parameters used in this work.  \label{tab:inputs}}
\end{center}
\end{table}

\section{\texorpdfstring{$X(x_t)$}{X(xt)} and \texorpdfstring{$X_c$}{Xc} expressions}\label{app:Xxt}
The short-distance contribution $X(x_t)$ in the SM (extracted from the
original papers~\cite{Buchalla:1993bv, Misiak:1999yg, Buchalla:1998ba,Brod:2010hi}) 
is given in~\cite{Buras:2015qea}
\begin{align}
X(x_t) = X_0(x_t) + \frac{\alpha_s(\mu_t)}{4\pi}X_1(x_t) + \frac{\alpha}{4\pi}X_{\rm EW}(x_t),
\end{align}
where $X_0$ is the leading order result, 
and $X_1$, $X_{\rm EW}$ are the NLO QCD and EW corrections, respectively. 
The coupling constants $\alpha_s$ and $\alpha$, as well as the parameter 
$x_t = m_t^2/m_W^2$, 
have to be evaluated at scale $\mu\sim\mathcal{O}(M_t)$.
The LO expression is the
gauge-independent linear combination $X_0(x_t) \equiv C(x_t) - 4 B(x_t)$~\cite{Inami:1980fz,Buchalla:1990qz}
\begin{equation}\label{X01}
X_0(x_t) = \frac{x_t}{8}\left[\frac{x_t+2}{x_t-1} + \frac{3x_t-6}{(x_t-1)^2}\log x_t\right].
\end{equation}
The NLO QCD correction \cite{Buchalla:1993bv,Misiak:1999yg,Buchalla:1998ba}, in the $\overline{\text{MS}}$ scheme reads
\begin{equation}\begin{aligned}
X_1(x_t) &= -\frac{29x_t - x_t^2 - 4x_t^3}{3(1-x_t)^2} - \frac{x_t + 9x_t^2 - x_t^3 - x_t^4}{(1-x_t)^3}\log x_t\\
&+ \frac{8x_t + 4x_t^2 + x_t^3 - x_t^4}{2(1-x_t)^3}\log^2 x_t - \frac{4x_t - x_t^3}{(1-x_t)^2}{\rm Li}_2(1-x_t)\\
&+ 8x_t\frac{\partial X_0}{\partial x_t}\log\frac{\mu^2}{M_W^2},
\end{aligned}\end{equation}
where $\mu$ is the renormalisation scale. The 2-loop EW correction $X_{\rm EW}$ has been calculated in \cite{Brod:2010hi}.
The charm contributions, $X_c^\nu(\equiv \lambda^4 P_c(X))$, are described via
\begin{align}
 P_c(X)= P_c^{\rm SD}(X) + \delta P_{c,u}
\end{align}
where $\delta P_{c,u}=0.04\pm0.02$ corresponds to the long-distance contributions as calculated in ref.~\cite{Isidori:2003ts}.
The short-distance contribution of the charm quark $P_c(X)$ including  NNLO correction is calculated in ref.~\cite{Brod:2008ss}
but the explicit analytical expression is not given. However, an approximate formula is given by\footnote{
The approximate formula in ref.~\cite{Brod:2008ss} is given for $\lambda=0.2255$, 
to take into account changes of $\lambda$, it should be multiplied by $(\lambda/0.2255)^4$.
}
\begin{equation}\label{eq:PCSD}
  \begin{split}
P_c^{\rm SD}(X) &= 0.38049
\left( \frac{m_c (m_c)}{1.30 \textrm{GeV}} \right)^{0.5081}
\left( \frac{\alpha_s (M_Z)}{0.1176} \right )^{1.0192} 
\left( 1 +
  \sum_{i,j} \kappa_{ij} L_{m_c}^i L_{\alpha_s}^j
\right) \\
&\pm 0.008707 
\left( \frac{m_c (m_c)}{1.30 \textrm{GeV}} \right)^{0.5276}
\left( \frac{\alpha_s (M_Z)}{0.1176} \right )^{1.8970} 
\left( 1 +
  \sum_{i,j} \epsilon_{ij} L_{m_c}^i L_{\alpha_s}^j
\right) \, ,       
  \end{split}
\end{equation}
where 
\begin{equation} \label{eq:defLs}
L_{m_c} = \ln \left( \frac{m_c (m_c)}{1.30 \textrm{GeV}} \right) \, , 
\qquad
L_{\alpha_s} = \ln \left( \frac{\alpha_s (M_Z)}{0.1176} \right) \, ,
\end{equation}
and 
\begin{align}
 &\kappa_{10} = 1.6624,\quad \kappa_{01} = -2.3537,\quad \kappa_{11} = -1.5862,\quad \kappa_{20} =  1.5036,\quad \kappa_{02} =  -4.3477,\nonumber\\
 &\epsilon_{10} = -0.3537,\quad \epsilon_{01} =  0.6003,\quad\epsilon_{11} =  -4.7652,\quad\epsilon_{20} = 1.0253,\quad\epsilon_{02} =  0.8866.
\end{align}

The NP effects that are neutrino-flavour dependent, beside NP$\times$NP terms contribute via SM$\times$NP 
interference terms.
Thus, for these types of NP effects in ${\rm BR}(K^+ \to \pi^+ \nu \bar{\nu})$ we need the NNLO charm 
contributions for the different neutrino flavours in a separated form (see eq.~\ref{eq:Br-KppipnunuExpanded} below).
However, since the charm contributions are not available separately at NNLO as given in eq.~\ref{eq:PCSD}, 
for the SM$\times$NP interference terms we use
the NLO results from appendix~C.2 of ref.~\cite{Bobeth:2016llm} which is given for $\mu_c = 1.3$ GeV
\begin{align}\label{eq:Xcflavour}
 X_c^{e/\mu}&= 10.05\times 10^{-4}, &&  X_c^{\tau}= 6.64\times 10^{-4}.
\end{align}

The SM$\times$NP interference terms are clearly seen when having eqs.~\ref{eq:Br-KLpinunu} and~\ref{eq:Br-Kppipnunu} in their expanded form 
\begin{align}
  \label{eq:Br-KLpinunuExpanded}
  & {\rm BR}(K_L \to \pi^0 \nu \bar{\nu})  =  \frac{\kappa_L }{\lambda^{10}}\frac{1}{3}s_W^4 \sum_{\nu_\ell}
  {\rm Im}^2 \left[\lambda_t C_L^{\nu_\ell} \right] \\[4pt]\nonumber
  &=  {\rm BR}(K_L \to \pi^0 \nu \bar{\nu})_{\rm SM} 
   + \frac{\kappa_L }{\lambda^{10}}\frac{1}{3} s_W^4 \Bigg[ \sum_{\nu_\ell} \mbox{Im}^2\left( \lambda_t\, C_{L,{\rm NP}}^{\nu_\ell} \right) 
  +2\,\mbox{Im}\left(\lambda_t\, C_{L,{\rm SM}} \right)
  \sum_{\nu_\ell} \mbox{Im}\left(\lambda_t \,C_{L,{\rm NP}}^{\nu_\ell} \right) \Bigg] \\[20pt]
%
\label{eq:Br-KppipnunuExpanded}
  &{\rm BR}(K^+ \to \pi^+ \nu \bar{\nu})  =
  \frac{\kappa_+ (1 + \Delta_{\rm EM})}{\lambda^{10}}\frac{1}{3} s_W^4 \sum_{\nu_\ell}
  \left[  {\rm Im}^2 \Big(\lambda_t C_L^{\nu_\ell} \Big)
        + {\rm Re}^2 \Big(-\frac{\lambda_c X_{c}}{s_W^2}
                         + \lambda_t C_L^{\nu_\ell} \Big)\right]\\[4pt]\nonumber
 &={\rm BR}(K^+ \to \pi^+ \nu \bar{\nu})_{\rm SM} 
   + \frac{\kappa_+ (1 + \Delta_{\rm EM})}{\lambda^{10}}\frac{1}{3} s_W^4 
  \Bigg[ \sum_{\nu_\ell} \mbox{Im}^2\left( \lambda_t C_{L,{\rm NP}}^{\nu_\ell} \right) 
 + 2\,\mbox{Im}\left(\lambda_t C_{L,{\rm SM}} \right)\sum_{\nu_\ell} \mbox{Im}\left(\lambda_t C_{L,{\rm NP}}^{\nu_\ell} \right) \\[-6pt]\nonumber 
     &\quad+ \sum_{\nu_\ell} \mbox{Re}^2\left( \lambda_t C_{L,{\rm NP}}^{\nu_\ell} \right) 
  +2\,\mbox{Re}\left(\lambda_t C_{L,{\rm SM}} \right)\sum_{\nu_\ell} \mbox{Re}\left(\lambda_t C_{L,{\rm NP}}^{\nu_\ell} \right)
   -2\sum_{\nu_\ell} \mbox{Re}\left(\frac{\lambda_c X_c^\nu}{s_W^2} \right) \mbox{Re}\left(\lambda_t C_{L,{\rm NP}}^{\nu_\ell} \right) \Bigg]\nonumber
\end{align}

\section{Results with \texorpdfstring{$\delta C_L^e=\delta C_L^\tau $}{CLe=CLtau}}
\label{app:otherpossibility}
In this section, we provide the results of the scan corresponding to the  possibility where the NP Wilson coefficients for the electron and tau are set equal to each other: $\delta C_L^e=\delta C_L^\tau $. As tau contribution is relevant only for the decays involving the neutrinos, we present the other possibility for figures~\ref{fig:Kppinunu_CLeCLmu} and~\ref{fig:all_obs_individually}. 
\begin{figure}[t!]
\begin{center}
\includegraphics[width=0.48\textwidth]{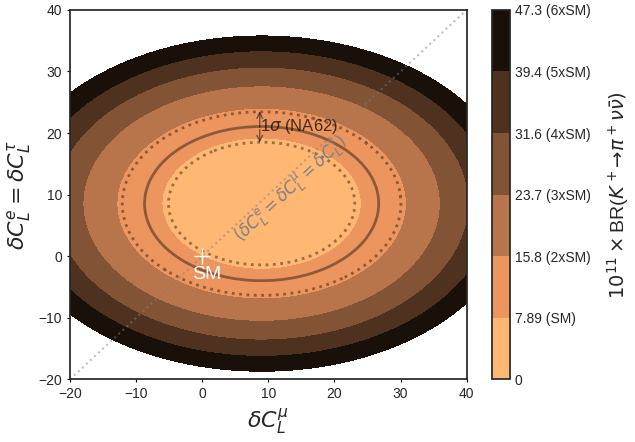}\quad  \includegraphics[width=0.49\textwidth]{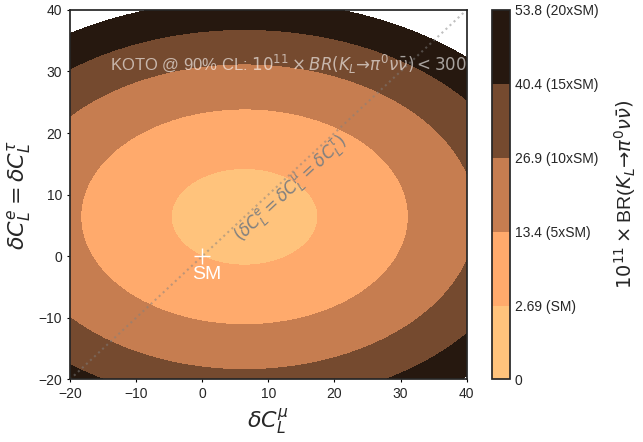}
\caption{BR($K^+\to \pi^+ \nu \bar{\nu}$) (left) and 
BR($K_L\to \pi^0 \nu \bar{\nu}$) (right) as a function of 
$\delta C_L^e=\delta C_L^\tau$ and $\delta C_L^\mu$. In the left plot, the solid  (dashed) line corresponds to the measured central value ($1\sigma$ experimental uncertainty) by NA62~\cite{NA62:2021zjw}. In the right the upper bound on BR($K_L\to \pi^0 \nu \bar{\nu}$) is not visible for the values scanned.
\label{fig:Kppinunu_CLeCLmub}}
\end{center}
\end{figure}
\begin{figure}[t!]
\begin{center}
\includegraphics[width=0.48\textwidth]{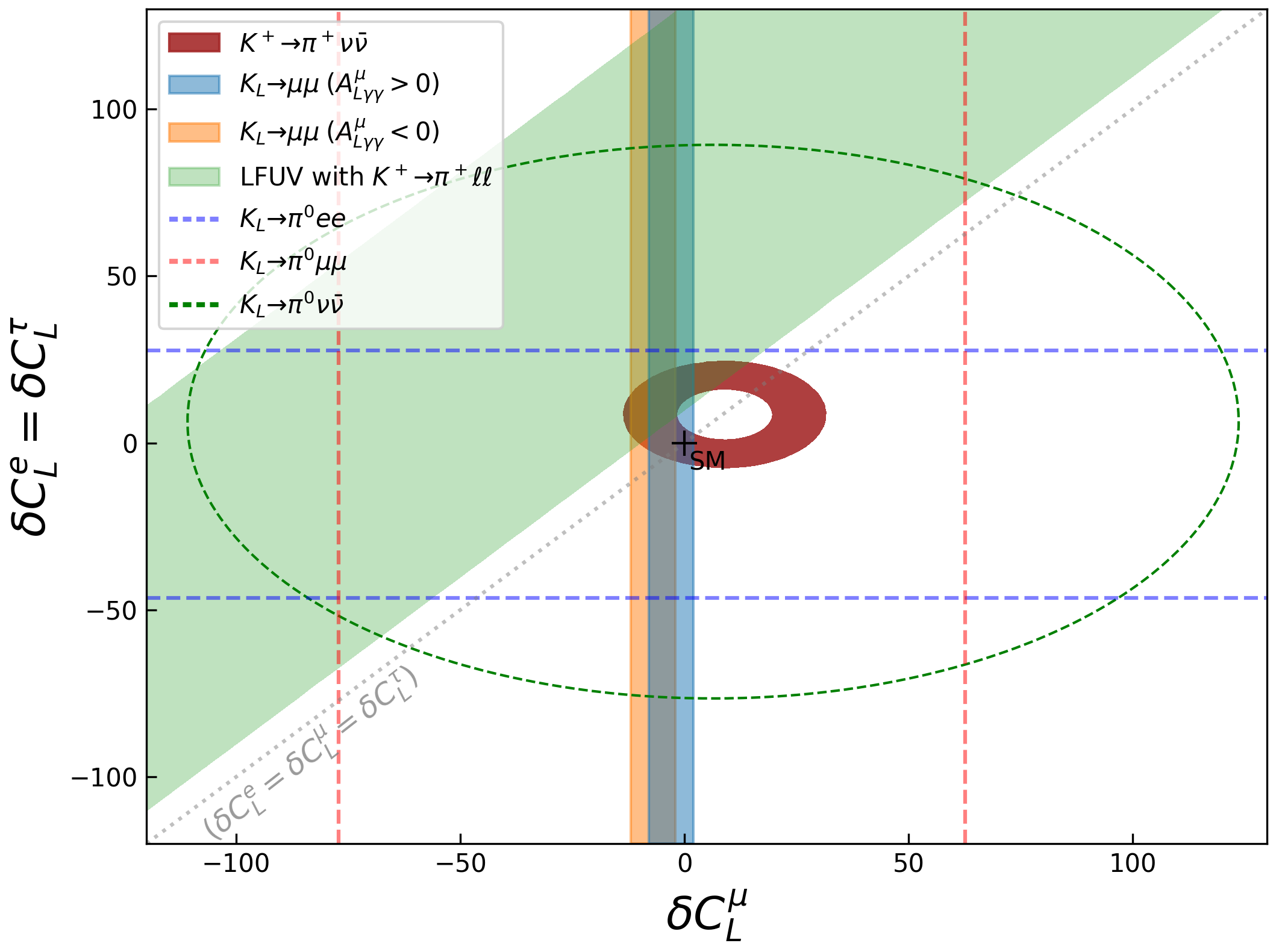}
\includegraphics[width=0.48\textwidth]{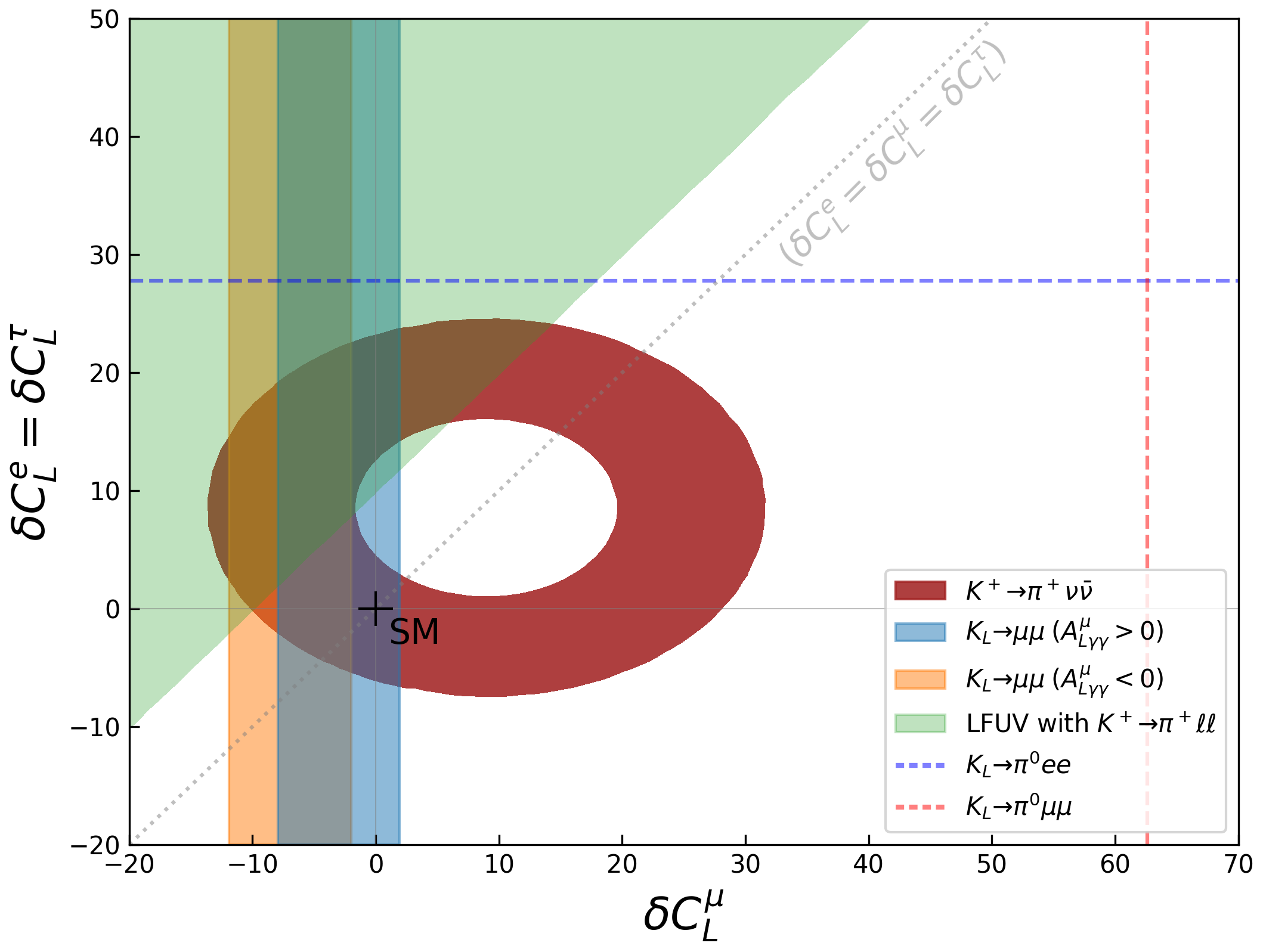}
\caption{\small
The bounds from individual observables. The right panel is the zoomed version of the left panel. The coloured regions correspond to  $68\%$ CL when there is a measurement and the dashed ones to upper limits.
$K_L \to \mu  \bar{\mu}$ has been shown for both signs for the long-distance contribution. For $K_L \to \pi^0 e \bar{e}$ and $K_L \to \pi^0 \mu \bar{\mu}$, constructive interference
between  direct  and  indirect  CP-violating  contributions  have  been  assumed. 
\label{fig:all_obs_individuallyb}}
\end{center}
\end{figure}
In the case of the former,
figure~\ref{fig:Kppinunu_CLeCLmub} illustrates corresponding regions when the New Physics Wilson coefficients for the electron and tau are set equal to each other. This has implications when we present the combination of all observables and is shown in figure~\ref{fig:all_obs_individuallyb}. In comparison with figure~\ref{fig:all_obs_individually}, note the flattening of the maroon ellipse which makes the upper bound for $K_L\to\pi^0e \bar{e}$ less powerful in constraining the regions of the parameter space that were scanned.


\section{\texorpdfstring{$Y(x_t)$}{Y(xt)} and \texorpdfstring{$Y_c$}{Yc} expressions}\label{app:Yxt}
The SM expressions of  $Y(x_t)$ is given in ref.~\cite{Buchalla:1998ba} 
\begin{align}\label{eq:Yfull}
Y(x_t) = Y_0(x_t) + \frac{\alpha_s(\mu_t)}{4\pi} Y_1(x_t),
\end{align}
where the LO gauge-independent $Y_0(x_t)\equiv C_0(x_t)-B_0(x_t)$ 
\begin{align}
Y_0(x) = \frac{x_t}{8}\left[\frac{4-x_t}{1-x_t}+\frac{3x_t}{(1-x_t)^2}\log x_t\right]
\end{align}
and 
\begin{align}
Y_1(x_{t}) &= \frac{10x_{t} + 10 x_{t}^2 + 4x_{t}^3}{3(1-x_{t})^2} -
           \frac{2x_{t} - 8x_{t}^2-x_{t}^3-x_{t}^4}{(1-x_{t})^3} \log x_{t}\nonumber\\
         &+\frac{2x_{t} - 14x_{t}^2 + x_{t}^3 - x_{t}^4}{2(1-x_{t})^3} \log^2 x_{t}
           + \frac{2x_{t} + x_{t}^3}{(1-x_{t})^2} {\rm Li_2}(1-x_{t})\nonumber\\
         &+8x_{t} \frac{\partial Y_0}{\partial x_{t}} \log \frac{\mu^2}{M^2_W}.
\end{align}

For the charm contributions, we have $Y_c = \lambda^4 P_c(Y)$ where 
$P_c(Y)$ is calculated at NNLO in QCD~\cite{Gorbahn:2006bm}.
The analytic expression is not given, however, an approximate formula with $\lambda$-dependence is offered
\begin{align}\label{eq:PcYlambda}
 P_c(Y) & = 0.115 \pm 0.008_{\rm theor} \pm 0.008_{m_c} \pm 0.001_{\alpha_s} 
 = \left ( 0.115 \pm 0.018 \right ) \left ( \frac{0.225}{\lambda} \right)^4 \, ,     
\end{align}
Another approximate formula with an accuracy of better than $\pm1.0\%$ in the ranges
$1.15 \! \text{ GeV} \le m_c (m_c) \le 1.45 \! \text{ GeV}$, $0.1150 \le \alpha_s (M_Z)
\le 0.1230$, $1.0 \! \text{ GeV} \le \mu_c \le 3.0 \! \text{ GeV}$ and $2.5 \! \text{ GeV}
\le \mu_b \le 10.0 \! \text{ GeV}$ is  also given in ref.~\cite{Gorbahn:2006bm}
\begin{align}\label{eq:PcYscales}
 P_c(Y) & = 0.1198 \left ( \frac{m_c (m_c)}{1.30 \! \text{ GeV}} \right )^{2.3595}
\left ( \frac{\alpha_s (M_Z)}{0.1187} \right )^{6.6055} \nonumber\\
 &\times \left ( 1 + \sum_{i,j,k,l} \kappa_{ijlm} L_{m_c}^i L_{\alpha_s}^j
L_{\mu_c}^k L_{\mu_b}^l \right )\left(\frac{0.225}{\lambda} \right)^4 ,
\end{align}
where
\begin{align}
 L_{m_c} & = \ln \left ( \frac{m_c (m_c)}{1.30 \! \text{ GeV}} \right ) \, , & 
L_{\alpha_s} & = \ln \left ( \frac{\alpha_s (M_Z)}{0.1187} \right ) \,
, \nonumber\\
L_{\mu_c} & = \ln \left ( \frac{\mu_c}{1.5 \! \text{ GeV}} \right ) \, , & 
L_{\mu_b} & = \ln \left ( \frac{\mu_b}{5.0 \! \text{ GeV}} \right )
\, ,
\end{align}
with
\begin{align}
\kappa_{1000} &= -0.5373, && \kappa_{0100} = -6.0472, && \kappa_{0010} = -0.0956, \nonumber\\[-2mm]
\kappa_{0001} &= 0.0114, && \kappa_{1100} = 3.9957, && \kappa_{1010} = 0.3604, \nonumber\\[-2mm]
\kappa_{0110} &= 0.0516, && \kappa_{0101} = -0.0658, && \kappa_{2000} = -0.1767, \nonumber\\[-2mm]
\kappa_{0200} &= 16.4465, && \kappa_{0020} = -0.1294, && \kappa_{0030} = 0.0725. 
\end{align}

\section{The long-distance contribution to \texorpdfstring{$K_L \to \mu \bar{\mu}$}{KL -> mu mu}}\label{app:NLLD}
The long-distance contributions given in eq.~\ref{eq:LDKmumu} can be written as~\cite{DAmbrosio:2017klp,Chobanova:2017rkj}
\begin{align}\label{eq:NSLD}
N_{L}^{\rm LD} & = \frac{\pm4\, \alpha_{\rm em}\, m_\mu}{\pi\, f_K\, M_K^2\, }
\sqrt{\frac{2\pi}{M_K}\frac{{\rm Br}(K_L^0 \to \gamma \gamma)^{\rm EXP}}{\tau_L}}  (\chi_{\rm disp} + i\chi_{\rm abs})\,,
\end{align}
with ${\rm Br}(K_L^0 \to \gamma \gamma)^{\rm EXP} = (5.47 \pm 0.04)\times 10^{-4}$~\cite{PDG2020},
and $(\chi_{\rm disp} + i\chi_{\rm abs}) $ corresponding to the $2\gamma$ intermediate state given by~\cite{DAmbrosio:1986zin,GomezDumm:1998gw,Knecht:1999gb,Isidori:2003ts}
\begin{align}
\chi_{\rm abs} & = {\rm Im}\left(C_{\gamma\gamma}\right) = \frac{\pi}{2\beta_{\mu,K}}\ln\left( \frac{1-\beta_{\mu,K}}{1+\beta_{\mu,K}} \right)\,,\\
\chi_{\rm disp} &= \chi_{\gamma \gamma}(\mu) - \frac{5}{2} +\frac{3}{2}\ln\left(\frac{m_\mu^2}{\mu^2}\right) + {\rm Re}\left( C_{\gamma\gamma} \right)\,,
\end{align}
with
\begin{align}
\beta_{\mu,K} &= \sqrt{1-4m_\mu/M_K^2}\,, \\ 
C_{\gamma\gamma} &= \frac{1}{\beta_{\mu,K}}\left[ {\rm Li}_2\left( \frac{\beta_\mu -1}{\beta_\mu +1} \right) + \frac{\pi^2}{3} + \frac{1}{4}\ln^2 \left(\frac{\beta_\mu -1}{\beta_\mu +1}  \right)\right]\,,
\end{align}
where the low-energy coupling $\chi_{\gamma \gamma}(\mu)$ which depends on the $K_L \to \gamma \gamma$ form factor behaviour outside the physical region is estimated in ref.~\cite{Isidori:2003ts}
\begin{align}
 \chi_{\gamma \gamma}(M_\rho) = 5.83 \pm 0.15_{\rm exp} \pm 1.0_{\rm th}\,,
\end{align}
resulting in
$(\chi_{\rm disp} + i\chi_{\rm abs}) = (0.71 \pm 0.15 \pm 1.0) +i(-5.21)$~\cite{Mescia:2006jd}.

\section{Calculations of theory error}\label{app:error}
The theoretical errors for $K_L\to\mu \bar{\mu}$ and $K_L\to\pi^0\ell \bar{\ell}$ are characterised by  asymmetric uncertainties. In particular, for the former, the degree of departure from the symmetric Gaussian errors was significant for the  former and in contrast with the values quoted thus far.
In this section, we elaborate on the reasons for this departure and argue why a symmetric Gaussian uncertainty does not accurately reflect the true theoretical uncertainty.

\begin{figure}[t!]
\begin{center}
\includegraphics[width=0.57\textwidth]{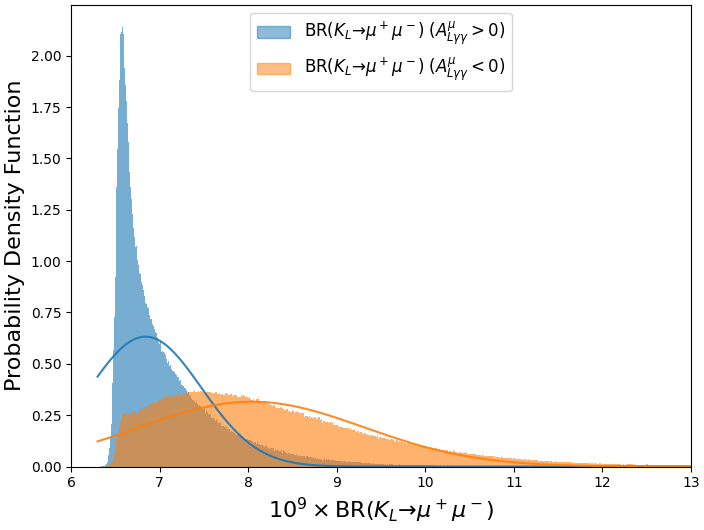}
\caption{\small
Comparison of the Monte Carlo PDF for $K_L\to\mu \bar{\mu}$ with the Gaussian description of the uncertainty.
\label{fig:error1}}
\end{center}
\end{figure}

For any given observable, we take into account the latest values of the inputs and the corresponding errors assuming a Gaussian distribution and employing a Monte Carlo simulation for the uncertainty propagation.  The blue (orange) distribution in figure~\ref{fig:error1} gives the probability distribution function (PDF)  for the positive (negative) sign of the long-distance contribution for $K_L\to\mu \bar{\mu}$.
The central values quoted in table~\ref{tab:data} reflect the value estimated by using the central values of the inputs given in table~\ref{tab:inputs}, while the asymmetric uncertainties are calculated by considering the boundaries between which the area under the PDF curve  is $0.68$.  This is to be compared with the  solid lines which reflect the   Gaussian description where $\mu$ and $\sigma$ have been naively calculated from the Monte Carlo distribution without taking into account the asymmetric nature of the distribution. We emphasise the large departure of the symmetric description compared to the Monte Carlo distribution, especially for the case of positive sign for long-distance contributions ($A_{L\gamma\gamma}^\mu>0$, shown in blue).

\begin{figure}[t!]
\begin{center}
\includegraphics[width=0.48\textwidth]{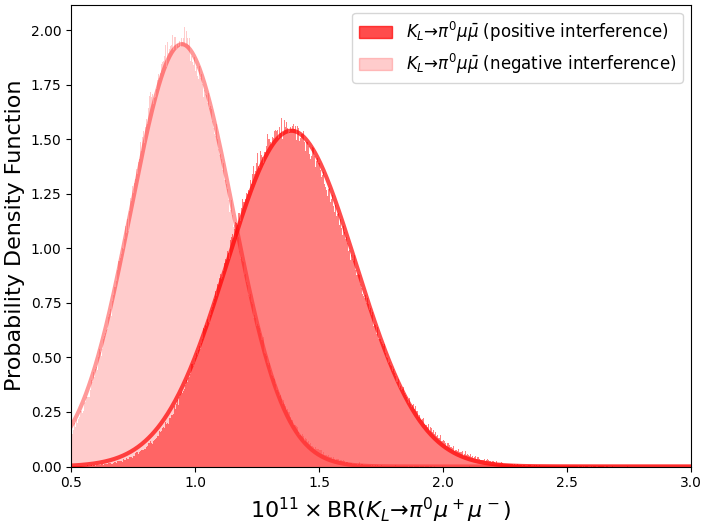}
\includegraphics[width=0.48\textwidth]{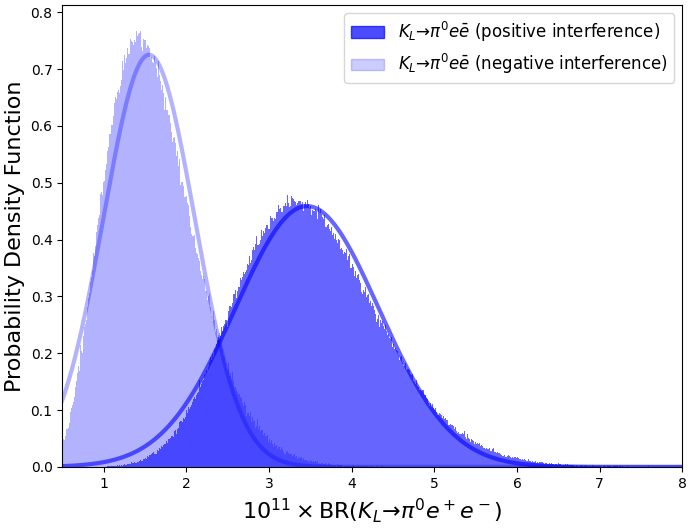}
\caption{\small
Comparison of the Monte Carlo PDF for $K_L\to\pi^0\ell \bar{\ell}$ with the Gaussian description of the uncertainty for both positive and negative interference.
\label{fig:error2}}
\end{center}
\end{figure}

A similar situation, albeit to a much lesser degree, is also noted for $K_L\to\pi^0\ell \bar{\ell}$ as shown in figure~\ref{fig:error2}.
For completeness, we show the distributions for both the signs of the interference between the direct and indirect CP-violating terms.
The difference between the actual distribution and the corresponding naive Gaussian description is only marginal. The values in table~\ref{tab:data} reflect the values obtained from the actual asymmetric distribution.



\end{document}